\documentclass{jfm}
\usepackage{graphicx}
\usepackage{epstopdf, epsfig}
\usepackage{xcolor}
\usepackage{pdfpages}
\usepackage{authblk}

\usepackage{amssymb}
\usepackage{amsmath}
\usepackage{txfonts}
\usepackage{mathdots}
\usepackage[classicReIm]{kpfonts}

\newtheorem{theorem}{Theorem}

\definecolor{codegreen}{rgb}{0,0.6,0}
\definecolor{codegray}{rgb}{0.5,0.5,0.5}
\definecolor{codepurple}{rgb}{0.58,0,0.82}
\definecolor{backcolour}{rgb}{0.95,0.95,0.92}

\shorttitle{Review Article}
\shortauthor{Naveen Balaji, Sujan Kumar S,Vignesh T,  K N Seetharamu}

\title{The Leading Edge Problem in Fluid Mechanics}

\author[1]{U S Naveen Balaji}
\author[2]{Sujan Kumar S}
\author[3]{T Vignesh}
\author[4]{Kankanhally N Seetharamu}
\author[5]{T R Seetharam}
\author[6]{Babu Rao Ponangi}
\author[7]{Rammohan B}

\affil[1,2,3]{PES University,Bangalore -
560085, Karnataka, India, naveenandmetallica@gmail.com}
\affil[4,5]{Chair Professor, Department of Mechanical Engineering, PES University, Bangalore -
560085, Karnataka, India.}
\affil[6]{Assistant Professor, Department of Mechanical Engineering, PES University, Bangalore - 560085, Karnataka, India}
\affil[7]{Associate Professor, Department of Mechanical Engineering, PES University, Bangalore - 560085, Karnataka, India}
\begin{document}

\maketitle
 
\begin{abstract}
The self-similar momentum ordinary differential equation (MODE) and the self-similar partial differential equation (MPDE) have been derived and the investigation of the integrability of the MODE and the MPDE  has been done by performing Painlev\'e test. A detailed discussion of the leading order behavior of the MODE and the MPDE has been presented with the latter being analyzed for the cases in which terms of increasing orders of Reynolds number have been considered. We have provided a brief introduction to Lie point symmetries and have found the Lie infinitesimal operator which when acts on the MPDE to order $\mathcal{O}(R)$ satisfies the Lie symmetry condition. Explicit calculations and expressions for the Lie prolongation terms have been presented. We have also investigated the integrability of various self-similar equations that arise from the generalized self-similar equation for different values of constants $\alpha_{1,2,3}$. Foundational work on transitional boundary solutions has been presented and transition solutions have been found via application of a junction condition at the leading edge-trailing edge boundary domain. A detailed discussion of semi-analytical solutions via the homotopy perturbation method is presented. We find semi-analytical solutions to the Falkner-Skan equation and the MODE by considering a Taylor series expansion as the initial approximation. An algorithmic scheme that involves consideration of a multi-dimensional Taylor expansion as the initial approximation to the MPDE has been presented.
\end{abstract}


\section{The Leading Edge Problem: An Introduction}
\noindent The boundary layer solution as derived by \citet{Blasius} hold for large Reynolds numbers but becomes inaccurate at the leading edge \citet{vandyke}. Thus, it is necessary to obtain a consistent solution at the leading edge in order to better understand the origin of the Blasius solution. Boundary layer flow over a flat plate is a classical fluid mechanics problem that has received significant attention over the past years. A comprehensive collection of boundary layer theory is presented in \cite{Boundarytheory}. Prandtl's boundary theory provides a leading order solution to the asymptotic expansion of the Navier-Stokes equations for large Reynolds numbers. Several papers have tried to extend Blasius' solution to the leading edge. \cite{Alden} extended to higher-order terms and included the pressure distribution, which is neglected in the classical boundary layer theory. \cite{Kuo} used a technique proposed by \cite{Lighthill} by which the solution of an approximated non-linear equation can be extended in the neighbourhood of a singularity merely by straining the argument of the solution to improve Blasius' solution such that it's validity is extended up to the leading edge. Imai \cite{Imai} was able to derive a first-order solution to Prandtl's boundary layer theory and showed that the vorticity decays exponentially with distance from the plate. \cite{Avramenko} investigated self-similar boundary layers in nanofluids and found the solutions to be valid far downstream from the leading edge. A detailed description of the boundary layer theory along mathematical studies and their importance to the nonlinear theory of viscous flows is provided in \cite{Oleinik}. We have closely followed [\cite{integrability1}, \cite{integrability2}] for theory on integrability and integrability tests.

\section{Momentum Equation: Problem Definition}\label{sec:momsection}
\subsection{Governing Equations and Boundary Conditions}\label{sec:goveq}
\noindent We consider an incompressible Newtonian fluid flowing with a uniform velocity $U$ into the leading edge of a semi-infinite flat plate. The fluid obeys the continuity and the momentum equations, given by
\\
\begin{equation}\label{conti}
    \frac{\partial u}{\partial x} + \frac{\partial v}{\partial y} = 0,
\end{equation}
\\
\begin{equation}\label{mom}
    \begin{array}{lr}
    \large{\rho\left(u\frac{\partial u}{\partial x} +v\frac{\partial u}{\partial y} \right) = -\frac{\partial p}{\partial x} + \mu\left(\frac{\partial^{2} u}{\partial x^{2}} + \frac{\partial^{2} u}{\partial y^{2}} \right)},\\\\
    
    \large{\rho\left(u\frac{\partial v}{\partial x} +v\frac{\partial v}{\partial y} \right) = -\frac{\partial p}{\partial y} + \mu\left(\frac{\partial^{2} v}{\partial x^{2}} + \frac{\partial^{2} v}{\partial y^{2}} \right)},
    \end{array}
\end{equation}
\\
where $u$ and $v$ are the $x$ and $y$ components of velocities respectively, $p$ is pressure, $\rho$ is density, and $\mu$ is the viscosity. The boundary conditions are that at the plate $y=0$, the fluid obeys no slip and no penetration, i.e.,
\\
\begin{equation}\label{atplate}
    u=0,\ v=0.
\end{equation}
\\
Far-field conditions dictate that as $y\rightarrow \infty$, the fluid obeys the free-stream conditions, i.e.,
\\
\begin{equation}\label{farfield}
    u\rightarrow U,\ v\rightarrow 0.
\end{equation}
\\
At the leading edge $x=0$, a uniform stream is imposed as follows
\\
\begin{equation}\label{leadingedge}
    u=U,\ v=0.
\end{equation}
\\
\subsection{Streamfunction}\label{sec:stream}
\noindent We now introduce the stream function $\psi$ in order to restate the momentum equations in terms of one single variable. We define
\\
\begin{equation}
    u=\frac{\partial \psi}{\partial y},\ v=-\frac{\partial \psi}{\partial x}.
\end{equation}
\\
The continuity equation in \ref{conti} is satisfied identically. The momentum equations as given in \ref{mom} become, with the subscript denoting differentiation,
\\
\begin{equation}
    \begin{array}{lr}
    \rho\left(\psi_{y}\psi_{xy} - \psi_{x}\psi_{yy} \right) = -p_{x} +\mu\left(\psi_{xxy} + \psi_{yyy} \right),\\\\
    
    \rho\left(-\psi_{y}\psi_{xx} + \psi_{x}\psi_{xy} \right) = -p_{y} -\mu\left(\psi_{xxx} + \psi_{xyy} \right).
    \end{array}
\end{equation}
\\
We now cross-differentiate to eliminate pressure and obtain
\\
\begin{equation}\label{pressureeliminator}
\rho\left(\psi_{y}\nabla^{2}\psi_{x} - \psi_{x}\nabla^{2}\psi_{y} \right) = \mu\nabla^{4}\psi,    
\end{equation}
\\
where $\nabla$ is a two-dimensional vector-operator, $\nabla = \left<\partial_{x},\partial_{y}\right>$. The boundary conditions can also be expressed in terms of the stream function. At the plate $y=0$, and \ref{atplate} yields
\\
\begin{equation}
    \psi_{y}=0,\ \psi_{x}=0.
\end{equation}
\\
Far from the plate as $y\rightarrow\infty$, \ref{farfield} gives
\\
\begin{equation}
    \psi_{y}\rightarrow U,\ \psi_{x}\rightarrow 0.
\end{equation}
\\
At the leading edge $x=0$, \ref{leadingedge} becomes
\\
\begin{equation}
    \psi_{y}=U,\ \psi_{x}=0.
\end{equation}\label{leadingedge_stream}
\\

\subsection{Self-Similar Transformation}\label{sec:selfsimilar}
\noindent At the leading edge, viscous forces dominate in \ref{pressureeliminator} and a balance between the viscous terms gives $y\sim x$. Hence, we define a self-similar variable
\\
\begin{equation}\label{selfsimvariable}
    \eta = \frac{y}{x},
\end{equation}
\\
and a self-similar stream function which depends on the self-similar variable $\eta$
\\
\begin{equation}\label{selfsmfuncODE}
    f(\eta) = \frac{\psi}{Ux}.
\end{equation}
\\
Now, substituting \ref{selfsimvariable} and \ref{selfsmfuncODE} into \ref{pressureeliminator} transforms the equation into the following ODE
\\
\begin{equation}\label{momentumODE}
    \begin{array}{lr}
    \left(1+\eta^{2}\right)^{2}f_{\eta\eta\eta\eta} + 8\eta\left(1+\eta^{2}\right)f_{\eta\eta\eta} + 4\left(1+3\eta^{2} \right)f_{\eta\eta}\\\\
    +\left(2\eta ff_{\eta\eta} + \left(1+\eta^{2} \right)\left(ff_{\eta\eta}\right)_{\eta}\right)R = 0.
    \end{array}
\end{equation}
\\
Now, following the same procedure we consider a self-similar function that depends on both the self-similar variable $\eta$ and the Reynolds number $R$ 
\\
\begin{equation}\label{selfsmfunc}
    f(\eta,R) = \frac{\psi}{Ux}.
\end{equation}
\\
Substituting \ref{selfsimvariable} and \ref{selfsmfunc} into \ref{pressureeliminator} transforms the equation into the following PDE
\\
\begin{equation}\label{momentumfourthPDE}
    \begin{array}{lr}
    \left(1+\eta^{2}\right)^{2}f_{\eta\eta\eta\eta} + 8\eta\left(1+\eta^{2}\right)f_{\eta\eta\eta} + 4\left(1+3\eta^{2} \right)f_{\eta\eta}\\\\
    +\left(2\eta ff_{\eta\eta} + \left(1+\eta^{2} \right)\left(ff_{\eta\eta}\right)_{\eta} -4\left(1+3\eta^{2}\right)f_{\eta\eta R} -4\eta\left(1+\eta^{2}\right)f_{\eta\eta\eta R}\right)R\\\\
    +\left(2\eta\left(f_{R}f_{\eta\eta} - ff_{\eta\eta R}\right)- \left(1+\eta^{2} \right)\left(f_{\eta}f_{\eta\eta R} - f_{R}f_{\eta\eta\eta} \right) + 2\left(1+3\eta^{2}\right)f_{\eta\eta RR}\right)R^{2}\\\\
    +\left(2\eta\left(f_{\eta}f_{\eta RR}-f_{R}f_{\eta\eta R}\right) + ff_{\eta RR} -3f_{\eta}f_{RR} + 4f_{RRR}-4\eta f_{\eta RRR}\right)R^{3}\\\\
    +\left(f_{RRRR}+f_{R}f_{\eta RR} - f_{\eta}f_{RRR}\right)R^{4}=0,
    \end{array}
\end{equation}
\\
where $R=\rho Ux/\mu$ is the local Reynolds number. It is to be mentioned here that the expression given in \cite{Rao} is incorrect and the equation \ref{momentumfourthPDE} is the correct form of the self-similar momentum PDE (we have also pointed this out in \cite{Leading}. Note that \ref{momentumfourthPDE} is a fourth-order non-linear PDE and as we shall see, the solution to the PDE determines the characteristic behavior of the energy PDE and hence, the heat transfer coefficient. The boundary conditions can now be expressed in terms of the self-similar variables. We have discussed the ambiguities associated with obtaining boundary conditions in the case of the energy equation in future chapters since here, we have the self-similar stream function to depend on both the self-similar variable $\eta$ and $R$.

\section{Energy Equation: Problem Definition}
\noindent When $\psi=f(\eta,R)Ux$ and $T=T(\eta,R)$, where $R=Ux/\nu$, we obtain the following PDE
\\
\begin{equation}\label{energyeq}
    (1+\eta^{2})T_{\eta\eta}+2\eta T_{\eta}+R\left(Pr U-4\eta T_{\eta} \right)+R^{2}\left(T_{RR}-PrU\left(f_{\eta}f_{\eta R}-f_{R}f_{\eta \eta} \right) \right)=0,
\end{equation}
\\
where $Pr$ is the Prandtl number which we will set to unity for all our calculations. Notice that when $R\rightarrow 0$, \ref{energyeq} simplifies to the following
\begin{equation}
    2\eta T_{\eta}+\left(1+\eta^{2}\right)T_{\eta\eta}=0.
\end{equation}
 \\
Similarly, for $Pr\equiv 1$ and $R\in[0,\infty)$, \ref{energyeq} becomes
\\
\begin{equation}\label{Prenergyeq}
    (1+\eta^{2})T_{\eta\eta}+2\eta T_{\eta}+R\left(U-4\eta T_{\eta} \right)+R^{2}\left(T_{RR}-U\left(f_{\eta}f_{\eta R}-f_{R}f_{\eta \eta} \right) \right)=0.
\end{equation}
\\
We now write \ref{Prenergyeq} as follows
\\
\begin{equation}
    (1+\eta^{2})T_{\eta\eta}+R^{2}T_{RR}+2\eta(1-2R)T_{\eta}+G\left(f_{\eta},f_{R},f_{\eta R},f_{\eta\eta},R \right)=0,
\end{equation}
 \\
where
\\
\begin{equation}\label{energyPDE}
    G\left(f_{\eta},f_{R},f_{\eta R},f_{\eta\eta},R \right)=RU-R^{2}U\left(f_{\eta}f_{\eta R}-f_{R}f_{\eta \eta} \right).
\end{equation}
\\
Now, comparing \ref{Prenergyeq} with the standard form of a second-order PDE written as
\\
\begin{equation}
    Au_{xx}+Bu_{xy}+Cu_{yy}+Du_{x}+Eu_{y}+Fu+G=0,
\end{equation}
\\
we observe that $B=0$, $A=(1+\eta^{2})$, and $C=R^{2}$. Consider the following cases
\\\\
i.\ When $R\rightarrow 0$, $B^{2}-AC=0$ and \ref{Prenergyeq} becomes a parabolic PDE.\\\\
ii.\ At the plate $y=0$ which implies that $\eta=0$ and hence, $B^{2}-AC>0$ and \ref{Prenergyeq} becomes a hyperbolic PDE.\\\\
iii.\ Far away from the plate $y\rightarrow \infty$ which implies that $\eta \rightarrow \infty$ and hence, $B^{2}-AC>0$ and \ref{Prenergyeq} becomes a hyperbolic PDE.\\\\
iv.\ At the leading edge $x=0$ which implies that $\eta\rightarrow \infty$ and hence, $B^{2}-AC>0$ and \ref{Prenergyeq} becomes a hyperbolic PDE.\\
\\\\
Define a non-linear operator $N$ as follows
\\
\begin{equation}
    N:=\frac{1}{R^{2}}\left(1+\eta^{2}\right)\frac{\partial^{2}}{\partial\eta^{2}}+2\eta(1-2R)\frac{\partial}{\partial \eta},
\end{equation}
\\
such that
\\
\begin{equation}\label{hyperenergy}
    N[T]+T_{RR}=H,
\end{equation}
\\
in $\mathcal{V}_{R}$ where $H=G\left(f_{\eta},f_{R},f_{\eta R},f_{\eta\eta},R \right)$ and this PDE is defined in the open set $\mathcal{V}\in \mathbb{R}^{2}$. We can now define a hyperbolic initial boundary value problem as follows
\\
\begin{equation}
    \begin{array}{lr}
    N[T]+T_{RR}=H\ \ \ in\ \mathcal{V}_{R},\\
    T=T_{\infty}\ \ \ \ \ \ \ \ \ \ \ \ \ \ \ on\ \partial\mathcal{V}\times [0,R],\\
    T=h,\ T_{R}=g\ \ \ \ \ \ on\ \mathcal{V}\times \{R=0\},
    \end{array}
\end{equation}
\\
where $\mathcal{V}$ is an open set in $\mathbb{R}^{2}$, $\mathcal{V}_{R}=\mathcal{V}\times(0,R]$ for a fixed Reynolds number $R>0$, $H:\mathcal{V}_{R}\rightarrow \mathbb{R}$ and $h,g:\mathcal{V}\rightarrow \mathbb{R}$ are given and $T:\bar{\mathcal{V}}_{R}\rightarrow \mathbb{R}$ is unknown, $T=T(\eta,R)$. Here $H:\mathcal{V}_{R}\rightarrow \mathbb{R}$ is given and $N$ denotes, for each Reynolds number $R$ a second-order partial differential operator.

\subsection{Difficulty Associated with Solving the Energy Equation}\label{sec:energydifficult}
\noindent In order to solve for the heat transfer coefficient, we need to first find $f(\eta, R)$, i.e., the solution to the self-similar momentum equation. The momentum equation with the self-similar transformation is a fourth-order, non-linear equation which is subject to boundary conditions
\\
\begin{equation}\label{generalcondition1}
\begin{array}{lr}
    f_{\eta}=0,\\
    f+f_{R}R=0
\end{array}
\end{equation} 
\\
at the plate (i.e., $\eta=0$) and
\\
\begin{equation}\label{generalcondition2}
    \begin{array}{lr}
    f_{\eta}\rightarrow 1,\\
    f+f_{R}R\rightarrow \eta
    \end{array}
\end{equation}
\\
being the far-field conditions (i.e., at $\eta\rightarrow \infty$). These are the boundary conditions that are to be imposed in order to determine the constants of a general solution to the momentum PDE \ref{momentumfourthPDE}, in all future sections we refer to boundary conditions \ref{generalcondition1} and \ref{generalcondition2} as generalized boundary conditions. It will be shown that although the conditions are satisfied at the plate, the far-field conditions are not. Although in this case, the characteristic curve of the heat-transfer coefficient is similar to, except for a singularity at $\eta\rightarrow 0$, the one obtained by our calculations.\\
\noindent Now, since the leading edge is usually subjected to Reynold’s number of the order $10^{-3}$, we may ignore the terms of order $\mathcal{O}(R^{2})$ and above. This approximation yields the following PDE 
\\
\begin{equation}\label{momeqO1}
\begin{array}{lr}
    \left((1+\eta^{2})^{2}f_{\eta\eta\eta\eta}+8\eta(1+\eta^{2})f_{\eta\eta\eta} +4\eta(1+3\eta^{2})f_{\eta\eta}\right)\\
    + R\left(2\eta ff_{\eta\eta} +(1+\eta^{2})\left(ff_{\eta\eta}\right)_{\eta}-4(1+3\eta^{2})f_{\eta\eta R}-4\eta(1+\eta^{2})f_{\eta\eta\eta R}\right)=0.
    \end{array}
\end{equation}
\\
It is observed that the first part of the PDE, the part independent of $R$ can be solved by setting $u=f_{\eta\eta}$ in order to reduced the PDE to the following ODE
\\
\begin{equation}\label{momODE}
     \left((1+\eta^{2})^{2}u_{\eta\eta}+8\eta(1+\eta^{2})u_{\eta} +4\eta(1+3\eta^{2})u\right)=0,
\end{equation}
\\
whose solution reads
\\
\begin{equation}
    f(\eta) = \frac{2}{\pi}tan^{-1}(\eta) + \frac{2\eta}{\pi (1+\eta^{2})}.
\end{equation}
\\
The part of \ref{momeqO1} which involve terms of $\mathcal{O}(R)$ read
\\
\begin{equation}\label{momOR2}
    R\left(2\eta ff_{\eta\eta} +(1+\eta^{2})\left(ff_{\eta\eta}\right)_{\eta}-4(1+3\eta^{2})f_{\eta\eta R}-4\eta(1+\eta^{2})f_{\eta\eta\eta R}\right)=0.
\end{equation}
\\
This is a fourth-order non-linear PDE which is difficult to solve to obtain an analytic solution. The difficulty associated with solving \ref{momOR2} can be attributed to the undifferentiated term $f(\eta,R)$ in $\left(ff_{\eta\eta}\right)_{\eta}$ since if we were to make the substitution $u=f_{\eta\eta}$ as made previously, we would obtain the following PDE
\\
\begin{equation}
\begin{array}{lr}
    R(2\eta u\left(\int{\int{u\ d\eta}}\ d\eta\right) +(1+\eta^{2})\left(u_{\eta}\left(\int{\int{u\ d\eta}}\ d\eta\right)+u\int{u\ d\eta}\right)\\
    -4(1+3\eta^{2})u_{R}-4\eta(1+\eta^{2})u_{\eta R})=0.
    \end{array}
\end{equation}
\\
which is a non-linear partial-integro differential equation and this complicates the problem. One possible method could be to remove the partial differentiation of $f$ with respect to $R$ by assuming that Reynolds number depends on the self-similar variable $\eta$ as $R=\alpha \eta^{\beta}$, where $\alpha$ and $\beta$ are real numbers. With this, \ref{momOR2} becomes a fourth-order non-linear PDE in $\eta$ and has the following form
\\
\begin{equation}
    R\left(2\eta ff_{\eta\eta} +(1+\eta^{2})\left(ff_{\eta\eta}\right)_{\eta}-\frac{4(1+3\eta^{2})}{\alpha\beta\eta^{\beta-1}}f_{\eta\eta\eta}-\frac{4(1+\eta^{2})}{\alpha\beta\eta^{\beta-2}}f_{\eta\eta\eta\eta}\right)=0,
\end{equation}
\\
but this resulting PDE is still complicated and difficult to solve. We here propose an alternative method where we notice that the PDE \ref{momeqO1} has only one differential in $R$ and thus, we set $f_{\eta\eta\eta R} = \gamma f_{\eta\eta\eta}$ and $f_{\eta\eta R} = \gamma f_{\eta\eta}$, where $\gamma$ is a positive constant. Now, we rewrite the second part of the PDE which is dependent on $R$ as follows
\\
\begin{equation}
    R\left(2\eta ff_{\eta\eta}(1+\eta^{2})\left(ff_{\eta\eta}\right)_{\eta} -4\gamma (1+3\eta^{2})f_{\eta\eta} - 4\eta\gamma(1+\eta^{2})f_{\eta\eta\eta} \right) = 0,
\end{equation}
\\
which can now be split into
\\
\begin{equation}\label{PDE1}
    (1+\eta^{2})\left(ff_{\eta\eta} \right)_{\eta}-4\gamma(1+3\eta^{2})f_{\eta\eta}=\psi(R),
\end{equation}
\\
and
\\
\begin{equation}\label{PDE2}
     2\eta ff_{\eta\eta} - 4\eta\gamma(1+\eta^{2})f_{\eta\eta\eta}=\phi(R),
\end{equation}
\\
where $\phi(R)$ and $\psi(R)$ are arbitrary functions of the Reynolds number. We now substitute the expression for $f_{\eta\eta\eta}$ which can be obtained from \ref{PDE2} into \ref{PDE1} to obtain the following
\\
\begin{equation}
    \left((1+\eta^{2})f_{\eta}-4\gamma(1+3\eta^{2})\right)f_{\eta\eta}+\frac{1}{2\gamma}f^{2}f_{\eta\eta}- \frac{\psi(R)}{4\eta\gamma}f=\phi(R).
\end{equation}
\\
Now, setting $\phi(R)=0,\ \psi(R)=0,$ and $\gamma=1$, the PDE simplifies to the following
\\
\begin{equation}
    f_{\eta\eta}\left((1+\eta^{2})f_{\eta}-4(1+3\eta^{2})+\frac{1}{2}f^{2}\right) = 0,
\end{equation}
\\
which has the following solution
\\
\begin{equation}\label{momeq01}
    \left(f(\eta,R)\right)_{\mathcal{O}(R)} = \frac{  4\left(4 + 5\eta^{2}+3\eta^{4} + 
   3\eta (1 + \eta^{2})^{2} tan^{-1}(\eta) - 
   16\eta (1 + \eta^{2})^{2} C_{1}\right)}{\eta(5+3\eta^{2})+2(1+\eta^{2})^{2}tan^{-1}(\eta)-16(1+\eta^{2})^{2}C_{1}}.
\end{equation}
\\
\noindent To fix the constant $C_{1}$ we make use of the boundary condition at the plate, i.e., \ref{generalcondition1}. We find that the constant $C_{1}=\pm\left( 1/2\sqrt{2}\right)$. Now, the solution to the momentum PDE up to order $\mathcal{O}(R)$ would be a linear combination of the solutions of the first part of \ref{momeqO1} which is independent of $R$, i.e., $f(\eta)$ and the second part of \ref{momeqO1} which depends on $R$ up to order $\mathcal{O}(R)$, i.e., $\left(f(\eta,R) \right)_{\mathcal{O}(R)}$. Thus, the final solution for positive constants $\lambda$ and $\Omega$ read
\\
\begin{equation}\label{momsol}
    f(\eta,R)= \lambda f(\eta) + R\Omega\left(f(\eta,R) \right)_{\mathcal{O}(R)}.
\end{equation}
\\
Note that this is only an approximate solution to the momentum PDE where we have neglected terms of order $\mathcal{O}(R^{2})$. We can now use this result to solve the energy equation without having to compromise terms of order $\mathcal{O}(R^{2})$ although this too will only be an approximate solution due to the compromise made in the momentum PDE.

\section{Approximate Solution to the Energy Equation}\label{sec:approxenergy}
\noindent We here present a solution to the energy PDE without considering terms of order $\mathcal{O}(R^{2})$. Consider the energy PDE \ref{energyeq} with this mentioned approximation
\\
\begin{equation}
    R\left(PrU-4\eta T_{\eta}\right)+2\eta T_{\eta} +\left(1+\eta^{2} \right)T_{\eta\eta} = 0.
\end{equation}
\\
We find a general solution of the following form
\\
\begin{equation}
    T(\eta,R) = \int_{1}^{\eta}{\left(-PrUR\  _{2}F_{1}\left[\frac{1}{2},2R,\frac{3}{2},-K_{1}^{2}\right]K_{1}\left(1+K_{1}^{2}\right)^{2R-1}+\left(1+K_{1}^{2}\right)^{2R-1}C_{1}(R)\right)}+C_{2}(R),
\end{equation}
\\
where $_{2}F_{1}$ is the hypergeometric function, $K_{1}$ is an integration constant and $C_{1}(R)$ and $C_{2}(R)$ are functions of Reynolds number. This integral is immensely complicated to solve and hence we expand the hypergeometric series around $K_{1}=0$ to obtain
\\
\begin{equation}
    _{2}F_{1}\left[\frac{1}{2},2R,\frac{3}{2},-K_{1}^{2}\right] = 1-\frac{2}{3}RK_{1}^{2}+\frac{1}{5}\left(R+2R^{2}\right)K_{1}^{4} +\mathcal{O}\left((K_{1})^{6}\right).
\end{equation}
\\
Now, integrating this expansion, the solution reads
\\
\begin{equation}
\begin{array}{lr}
    T(\eta,R) \approx C_{2}(R) + 4^{-1+R}PrU\frac{(5+6R)\left(3+R\left(8+R\left(-5+2R\right)\right) \right)}{15\left(1+R\right)\left(1+2R\right)}\\\\
    \ \ \ \ \ \ \ \ \ \ - PrU\left(1+\eta^{2} \right)^{2R}\frac{\left(15 + R\left(58 + R\left(46-2\left(13+16R\right)\eta^{2} + 3\left(1+2R\right)^{2}\eta^{4}\right)\right) \right)}{60\left(1+R\right)\left(1+2R\right)}\\\\
    \ \ \ \ \ \ \ \ \ \ + PrU\left(\ _{2}F_{1}\left(\frac{1}{2},1-2R,\frac{3}{2},-1 \right) -\eta\ _{2}F_{1}\left(\frac{1}{2},1-2R,\frac{3}{2},-\eta^{2} \right) \right)C_{1}(R).
    \end{array}
\end{equation}
\\
Before the fluid flows past the plate, there is no heat transfer and hence the temperature is unaffected. Thus, we demand that $T(\eta=0,R) = T_{\infty}$ and hence, we observe that $C_{2}=T_{\infty}$. Now, $R\rightarrow 0$ can imply two things-either $x=0$ and hence $\eta\rightarrow \infty$ (which is the leading edge condition) or the velocity of the fluid is very small. If the plate is being maintained at a uniform temperature then the temperature at all points of the plate would not fluctuate when there is no fluid available for heat transfer. Thus, we obtain a condition $T(\eta\rightarrow\infty,R\rightarrow0) = T_{w}$ which is valid only in the leading edge. Note however that $\eta\rightarrow\infty$ is also a condition when we are far away from the plate, i.e., when $y\rightarrow\infty$ and at this position even when there is no fluid flowing the temperature is that of the free-stream. Thus, the far-field condition reads $T(\eta\rightarrow\infty,R\rightarrow 0) = T_{\infty}$. When there is a fluid flowing with some arbitrary velocity, then far away from the plate there would be gradients of temperature as a result of heat transfer between the plate and the flowing fluid. Here, we don't make any assumption regarding the characteristic of the gradients since we don't know the behavior of temperature with viscosity and a similar argument can be made at the plate when $y\rightarrow 0$ for arbitrary Reynolds number. Thus, in order to establish the boundary conditions at the plate and far-field for arbitrary Reynolds numbers, we are to consider temperature-viscosity models, each of which depends upon the type of fluid begin used. Assuming that we have an appropriate model that provides a relationship between the temperature and viscosity, there still exists an additional constraint-relationship between Reynolds number and the self-similar variable. In order to overcome this, we are to yet again resort to assuming Reynolds number models as done in the previous sections.\\

\noindent To demonstrate this let us assume the following dependence of temperature on it's variables $\eta$ and $R$
\begin{equation}
    T(\eta,R) = \alpha tan^{-1}(\eta^{\beta}) + \gamma tan^{-1}(R^{\epsilon}),
\end{equation}
where $\alpha,\beta,\gamma,$ and $\epsilon$ are positive constants. In this model, at the plate
\\
\begin{equation}
\begin{array}{lr}
    T(\eta\rightarrow 0, R)=\gamma tan^{-1}(R^{\epsilon}),\\ T(\eta\rightarrow 0, R\rightarrow 0) = 0\ \& \\
    T(\eta\rightarrow 0,R\rightarrow\infty) = \gamma\frac{\pi}{2}.
\end{array}
\end{equation}
At the leading edge, 
\\
\begin{equation}
\begin{array}{lr}
    T(\eta\rightarrow \infty, R)=\alpha \frac{\pi}{2}+\gamma tan^{-1}(R^{\epsilon}),\\
    T(\eta\rightarrow\infty, R\rightarrow 0) = 0\ \& \\
    T(\eta\rightarrow \infty,R\rightarrow\infty) = \alpha\frac{\pi}{2} + \gamma\frac{\pi}{2}.
\end{array}
\end{equation}
\\
As we had deduced earlier, far away from the plate, $T(\eta\rightarrow\infty, R\rightarrow 0) = T_{\infty}$ but in order to find $T(\eta\rightarrow \infty , R)$ and $T(\eta\rightarrow \infty,R\rightarrow\infty)$ we need to consider temperature-viscosity models such as the Seeton or the Wright model. Similarly, when there is heat transfer between the fluid and the plate, temperature gradients would develop which would be of the following form
\\
\begin{equation}
    \left(\frac{dT}{dy} \right)_{y=0} = -\frac{h(T_{w}-T_{\infty})}{k},
\end{equation}
\\
and since $T_{y} = xT_{\eta}$ for a fixed value of $x$, we have $T_{\eta}(\eta=0,R) = -\frac{hx(T_{w}-T_{\infty})}{k}$. Notice now that the value of $x$ can be fixed by our choice of $R$ and thus, at the plate we obtain the following variations of the temperature gradients for different choices of Reynolds number
\\
\begin{equation}
    \begin{array}{lr}
    T_{\eta}(\eta=0,R) = -\frac{hx(T_{w}-T_{\infty})}{k},\\
    T_{\eta}(\eta=0,R\rightarrow 0) = 0,\ \& \\
    T_{\eta}(\eta=0,R\rightarrow\infty) \rightarrow -\infty.
    \end{array}
\end{equation}
\\
There would also exist a gradient with respect to Reynolds number but this would be determined with the choice of the temperature-viscosity model chosen. For instance, if we choose that Seeton model, we have the following relation between the temperature and the kinematic viscosity
\\
\begin{equation}
    A+B\ ln(T) = ln\left(ln\left(\nu + 0.1 + e^{-\nu}K_{0}\left(\nu + 1.244067 \right) \right) \right),
\end{equation}
\\
where $A$ and $B$ are liquid-specific values and $K_{0}$ is the zero-order modified Bessel function of the second kind. We then obtain the following expression for $T$ as a function of $\nu$
\\
\begin{equation}
    T(\nu) = e^{-\frac{A}{B}} \left(ln\left(\nu + e^{-\nu}K_{0}\left(\nu + 1.244067 \right) \right) \right)^{\frac{1}{B}}.
\end{equation}
\\
Since Reynolds number depends on both the kinematic viscosity and the self-similar variable we can probe for separable solutions of the form
\\
\begin{equation}
    \bar{T}(R) = \Omega(\eta)T(\nu).
\end{equation}
\\
and hence for the separable solution $T(\eta,R)$ which reads
\\
\begin{equation}
    T(\eta,R) = \Phi(\lambda)\Lambda(R),
\end{equation}
\\
where $\Lambda(r) = \lambda \bar{T}(R)$ for some positive constant $\lambda$.

\section{Integrability Tests of the Self-Similar Equations}
\noindent The Painlev\'e test is carried out to predict whether a particular ODE (or PDE) or a system or ODE (or PDE) is (are) integrable. Here, we lay out the test procedure for the case of an ODE. Throughout this section, we follow the work of \cite{Polyanin-non-PDE}. Consider the following n$^{th}$-order ordinary differential equation,
\\
\begin{equation}\label{eq:PAinleveODE}
    \phi_{x}^{(n)} = F\left(x,\phi,\phi_{x}',\ldots,\phi_{x}^{(n-1)} \right),\ \phi_{x}^{(n)}\equiv \frac{d^{n}\phi}{dx^{n}}.
\end{equation}
\\
We now search for a solution to \ref{eq:PAinleveODE} in the form of a series with a movable pole type singularity, as follows
\\
\begin{equation}\label{eq:PainleveODEsol}
    \phi(x) = \frac{1}{\left(x-x_{0} \right)^{p}}\sum_{j=0}^{\infty}A_{j}\left(x-x_{0} \right)^{j},
\end{equation}
\\
where $x_{0}$ is an arbitrary number and $p$ is a positive integer. It is to be noted that the solution \ref{eq:PainleveODEsol} is to be general, and hence the coefficients $A_{j}$ must contain $n-1$ arbitrary constants. If there is more than one
solution \ref{eq:PainleveODEsol}, all of them must satisfy the above requirements. In the following section, we have performed the Painlev\'e test for the momentum ODE, explaining the steps involved in the test in a chronological fashion.

\subsection{Painlev\'e Test for Momentum ODE}\label{sec:PainleveODE}
\noindent We now perform the test for the momentum ODE \ref{momentumODE} to determine its integrability. To determine the leading term of the series \ref{eq:PainleveODEsol} which is characterized by the exponent $-p$ and the coefficient $A_{0}$,  we substitute the following single term
\\
\begin{equation}\label{PainlevemomODE}
    f=\frac{A_{0}}{\xi^{p}},\ \xi = \eta - \eta_{0},
\end{equation}
\\
into equation \ref{momentumODE} and multiply the resulting expression by $\xi^{p+4}$ to obtain,
\\
\begin{equation}\label{eq:PainmomODEsol}
    \begin{array}{lr}
    \left(1+\left(\xi + \eta_{0} \right)^{2} \right)^{2}(3+p)(2+p)(1+p)pA_{0} - 8(\xi + \eta_{0})\left(1+\left(\xi + \eta_{0} \right)^{2} \right)(2+p)(1+p)pA_{0}\xi\\\\
    + 4\left(1+3\left(\xi + \eta_{0} \right)^{2} \right)(1+p)pA_{0} \xi^{2} = R\left(2(1+p)pA_{0}^{2} \xi^{2-p} - 2\left(1+\left(\xi + \eta_{0} \right)^{2} \right)(1+p)pA_{0}^{2} \xi^{1-p} \right).
    \end{array}
\end{equation}
\\
As $\xi\to0$ for $p>0$, we notice that the only term on the RHS which makes a non-zero contribution is the second term multiplied to the Reynolds number $R$, i.e., $- 2R\left(1+\left(\xi + \eta_{0} \right)^{2} \right)(1+p)pA_{0}^{2} \xi^{1-p}$. Now, in order to prevent a blow up we must set $p=1$. It now follows from \ref{eq:PainmomODEsol} that 
\\
\begin{equation}
  \left(24 + 48\eta_{0}^{2} + 24\eta_{0}^{4} \right)A_{0} = 8R\left(1+\eta_{0}^{2}\right)A_{0}^{2},  
\end{equation}
\\
and hence, the leading term of the series is given by \ref{PainlevemomODE} with,
\\
\begin{equation}\label{eq:ODEcoefficient}
    p=1,\ A_{0} = \frac{3 \left(1+ \eta_{0}^{2} \right)}{R}.
\end{equation}
\\
Thus, the momentum ODE \ref{momentumODE} satisfies the first necessary condition of the Painlev\'e test, i.e., $p>0$. We now look for a series solution to \ref{momentumODE} of the following form
\\
\begin{equation}\label{eq:seriesmomODEPainleve}
    f =  \frac{3 \left(1+ \eta_{0}^{2} \right)}{R}\xi^{-1} + A_{1} + A_{2}\xi + A_{3}\xi^{2} + \ldots.
\end{equation}
\\
Substituting \ref{eq:seriesmomODEPainleve} into \ref{momentumODE}, collecting terms of like powers of $\xi$, and equating the coefficients to zero, we arrive at the following system of algebraic equations for $A_{j}$  
\\
\begin{equation}\label{eq:coefficientsmomODE}
    \begin{array}{lr}
    \xi^{0}:\ 4\eta_{0}\left(-1+2\eta_{0}^{2}+\eta_{0}^{4}\right) + R\left(1+\eta_{0}^{2}\right)A_{1}=0,\\\\
       
    \xi^{1}:\ -2 - 18 \eta_{0}^{2} - 4 R \eta_{0}A_{1} + RA_{2} + R\eta_{0}^{2} A_{2} = 0,\\\\
    
    \xi^{2}:\ 20\eta_{0} + 36\eta_{0}^{3}+ R\left(3 + 7\eta_{0}^{2}\right)A_{1} - 6R\eta_{1}\left(1 + \eta_{0}^{2}\right)A_{2} + 2R\left(1+2\eta_{0}^{2} + \eta_{0}^{4}\right)A_{3}= 0,\\\\

    \xi^{3}:\ 2 + 6 \eta_{0}^2 + R\eta_{0}A_{1} - R\left(1 + 3\eta_{0}^{2}\right)A_{2} + 4R\eta_{0}\left(1+\eta_{0}^{2}\right)A_{3} = 0,\\\\
    
    \xi^{4}:\ 6 \eta_{0} A_{2} + A_{3} - 3 \eta_{0}^2 A_{3} + 2 R \eta_{0} A_{1} A_{3} + R A_{2} A_{3} + R \eta_{0}^2 A_{2} A_{3} = 0,\\\\
    
    \xi^{5}:\ 3\eta_{0}A_{3}+ R\eta_{0}A_{2}A_{3} + RA_{3}^{2} + R\eta_{0}^{2}A_{3}^{2} = 0,\\\\
    
    \xi^{6}:\ 0\times A_{4} + 2R^{2}\eta_{0}A_{3}^{2} = 0.
    
    \end{array}
\end{equation}
\\
From these equations we fix the coefficients to be
\\
\begin{equation}\label{eq:PainleveODEcoeffiients}
    A_{0} = \frac{3\left(1+\eta_{0}^{2}\right)}{R},\ A_{1} =\frac{4\left(\eta_{0}-2\eta_{0}^{3}-\eta_{0}^{5}\right)}{R\left(1+\eta_{0}^{2}\right)^{2}},\ A_{2} = \frac{2\left(1+19\eta_{0}^{2}+3\eta_{0}^{4}+\eta_{0}^{6}\right)}{R\left(1+\eta_{0}^{2}\right)^{3}},\ A_{3} = 0.
\end{equation}
\\
We notice from the last equation in \ref{eq:coefficientsmomODE} that
\\
\begin{equation}
   0\times A_{4} + 2R^{2}\eta_{0}A_{3}^{2} = 0.
\end{equation}
\\
Since $A_{3}=0$, this equation is automatically satisfied for arbitrary $A_{4}$. We note here that the rest of the coefficients, i.e., $A_{5}, A_{5},\ldots$ can be determined successively from recurrence relations which take the form
\\
\begin{equation}\label{eq:multipliers}
    k_{j}A_{j} = \Lambda\left(A_{0},A_{1},A_{2},\ldots,A_{j-1} \right),\ j=1,2,\ldots.
\end{equation}
\\
To determine any $A_{j}$ the following equations must hold simultaneously for the solution,
\\
\begin{equation}\label{eq:conditions}
    k_{m} = 0,\ \Lambda\left(A_{0},A_{1},A_{2},\ldots,A_{j-1} \right)=0.
\end{equation}
\\
From the analysis carried out for the momentum ODE we observe that the multipliers $k_{m}$ of the coefficients $A_{m}$ are only determined by the leading terms of equation \ref{momentumODE}, which can be found via a leading order analysis. We rewrite the momentum ODE, \ref{momentumODE} as follows \cite{Integrability-Tabor}
\\
\begin{equation}
    f'''' = F\left(f''',f'',f',f,\eta \right),
\end{equation}
\\
where $F$ is analytic in $\eta$ and rational in its other arguments. Now, we substitute
\\
\begin{equation}
    f(\eta) = a\left(\eta - \eta_{0} \right)^{m},
\end{equation}
\\
with $m<1$ and obtain,
\\
\begin{equation}
\begin{array}{lr}
    8 a (-2 + m) (-1 + m) m \eta (-\eta_{0} + \eta)^{-3 + 
    m} \left(1 + \eta^2\right)\\\\
    + a (-3 + m) (-2 + m) (-1 + m) m (-\eta_{0} + \eta)^{-4 +m} \left(1 + \eta^2\right)^{2}\\\\
    + 4 a (-1 + m) m (-\eta_{0} + \eta)^{-2 + m} \left(1 + 3 \eta^2\right)\\\\
    + 2R a^{2} (-1 + m) m \eta (-\eta_{0} + \eta)^{-2 + 2 m} + R\left(1 + \eta^2\right)(a^{2} (-2 + m) (-1 + m) m (-\eta_{0} + \eta)^{-3 + 2 m}\\\\
    + Ra^{2} (-1 + m) m^{2} (-\eta_{0} + \eta)^{-3 + 2m}) = 0.
\end{array}
\end{equation}
\\
The most singular terms, i.e., those with the most negative exponents, called the \textit{dominant balance terms}, are to balance exponents and coefficients at the singularity. Here, the second term on the LHS  and the second and third terms on the RHS are the dominant terms. Hence,
\\
\begin{equation}
    -4+m=-3+2m\Rightarrow m=-1,
\end{equation}
\\
and,
\\
\begin{equation}
    24 a \left(1 + \eta^{2}\right)^{2} = 8 a^{2}R \left(1 + \eta^{2}\right)\Rightarrow a = \frac{3\left(1+\eta^{2} \right)}{R}.
\end{equation}
\\
The behavior of the solution is $f\sim \left(\eta - \eta_{0} \right)^{-1}$ and the behavior in the neighborhood of the singularity is given by expansion in a Laurent series, which reads
\\
\begin{equation}
    f(\eta) = \sum_{\beta=0}^{\infty}a_{\beta}\left(\eta - \eta_{0} \right)^{\beta - 1},
\end{equation}
\\
and plugging this back into the ODE yields recurrence relations which we will not write here. Upon, simplification, it can be seen that the right hand sides of equation \ref{eq:PainmomODEsol} are linear in $A_{m}$ while the left-hand sides are independent of $A_{m}$, the multipliers $k_{m}$ can be found by substituting the following two-term expression
\\
\begin{equation}\label{eq:binomial}
    f = A_{0}\xi^{-p} + A_{m}\xi^{m-p},\ \xi = \eta - \eta_{0}
\end{equation}
\\
into the leading terms of the momentum ODE. Upon doing this substitution, collecting terms the like powers of $A_{m}$ gives
\\
\begin{equation}
    A_{m}k_{m}\xi^{q} + \mathcal{O}\left(A_{m}^{2} \right) = 0,\ q\geq m-n-p,
\end{equation}
\\
where $k_{m}$ is the desired multiplier which can be represented as a polynomial of degree $n$, where $n$ is the order of the ODE ($n=4$ here), in the integer index $m$ as follows
\\
\begin{equation}
\begin{array}{lr}
    k_{m} = b_{n}m^{n} + b_{n-1}m^{n-1} + \ldots + b_{1}m + b_{0}.\\
    \ \ \ \ \ = b_{4}m^{4} + b_{3}m^{3} + b_{2}m^{2} + b_{1}m + b_{0}
\end{array}
\end{equation}
\\
The equation $k_{m} = 0$ always has a root $m = 1$ (which corresponds to the arbitrariness in the choice of $\eta_{0}$) and the other roots determine the so called \textit{Fuchs indices} which are also known as resonances, i.e., the numbers $m_{1}, \ldots ,m_{n-1}$ of the coefficients $A_{m}$ in the expansion \ref{eq:PainleveODEsol} that can be arbitrary. The second part of the Painlev\'e test consists of substitution of the binomial \ref{eq:binomial} into the leading terms of the momentum ODE and the multiplier $k_{m}$ appearing on the left-hand side of the relations in \ref{eq:multipliers} is determined as follows
\\
\begin{equation}\label{eq:Polynomialequation}
    k_{m}  = 24 - 50 m + 35 m^{2} - 10 m^{3} + m^{4}.
\end{equation}
\\
Now, from the expression $k_{m}=0$ we find the Fuchs indices $m_{1} = 1$, $m_{2} = 2$, $m_{3} = 3$, and $m_{4} = 4$. Thus, the Fuchs index is $m_{4}=4$ and thus, the second necessary condition of the Painlev\'e test is satisfied and four terms are to be considered in the expansion \ref{eq:seriessolmomODE} where the coefficient of the
last term is $A_{3}$ which can be arbitrary and hence, the $\xi^{2} = \left(\eta - \eta_{0} \right)^{2}$ is called the resonance or the \textit{Kovalevskaya exponent}. For the third and final condition, we substitute the expansion \ref{eq:seriesmomODEPainleve} into the momentum ODE as done in \ref{eq:coefficientsmomODE} and check whether conditions \ref{eq:conditions} hold simultaneously. Since this holds, we conclude that  the momentum ODE \ref{momentumODE} passes the Painlev\'e test with it's solution being expressed as the series expansion \ref{eq:seriesmomODEPainleve} with two arbitrary constants $\eta_{0}$ and $A_{3}$, i.e,
\\
\begin{equation}\label{eq:finalsolPainleveODE}
    f(\eta) = \frac{3\left(1+\eta_{0} \right)}{R}\xi^{-1} +\frac{4\left(\eta_{0}-2\eta_{0}^{3}-\eta_{0}^{5}\right)}{R\left(1+\eta_{0}^{2}\right)^{2}} + \frac{2\left(1+19\eta_{0}^{2}+3\eta_{0}^{4}+\eta_{0}^{6}\right)}{R\left(1+\eta_{0}^{2}\right)^{3}}\xi.
\end{equation}

\subsection{Painlev\'e Test for the Momentum PDE}\label{sec:PainlevePDE}
\noindent We probe for a solution in a small neighborhood of a manifold $\eta-\eta_{0}(R) = 0$ in the form of the following series expansion \cite{Jimbo-Kruskal-Miwa}
\\
\begin{equation}\label{momentumPDEPainleve}
    f(\eta,R) = \frac{1}{\xi^{p}}\sum_{m=0}^{\infty}f_{m}(R)\xi^{m},\ \xi(\eta,R) = \eta-\eta_{0}(R).
\end{equation}
\\
Here, the exponent $p$ is a positive integer, so that the movable singularity is of the pole type. The function $\eta_{0}(R)$
is assumed  to be arbitrary, and the $f_{m}$ are assumed to depend on derivatives of $\eta_0(R)$. We now substitute \ref{eq:seriesmomODEPainleve} into the momentum PDE \ref{momentumfourthPDE} and follow the given scheme \cite{Polyanin-non-PDE}:
\newline
\newline
1. Substitute into PDE, $f(\eta,R) = f_{0}\xi^{-p}$, $f_{0} = f_{0}(R)$, $\xi = \xi(\eta,R)$.
\newline
2. Find the constant $p$ and the function $f_{0}(\eta,R)$. Check if $p$ is positive (necessary condition).
\newline
3. Look for Fuchs indices, i.e., resonances.
\newline
4. Substitute $f = f_{0}\xi^{-p} + f_{m}\xi^{m-p}$ in the leading terms of the PDE and collect terms proportional to $f_{m}$.
\newline
5. Obtain the expression $k_{m}f_{m}\xi^{q}+\ldots$ where $k_{m}$ is a polynomial with respect to the integer index $m$.
\newline
6. Factorize the polynomial $k_{m}$ as $k_{m} = (m+1)(m-m_{1})\ldots(m-m_{n-1})$, where $n$ is the order of the PDE and $m_{j}$ are the resonances.
\newline
7. If $m_{1},\ldots,m_{n-1}$ are all non-negative, proceed with test. This is the first sufficient condition.
\newline
8. Obtain recurrence relations for expansion coefficients $f_{m}$ and check the consistency conditions. This is the second sufficient condition.
\newline
9. If all of the conditions are satisfied, the PDE passes the Painlev\'e test.

\subsubsection{Painlev\'e Test of PDE to Order $\mathcal{O}(R)$}
\noindent We first solve the PDE to order $\mathcal{O}(R)$ and have presented the results of the test following the above steps. Following step $1$ and substituting the leading term of the series into the entire momentum PDE, we obtain explicit expressions (which we won't write here as they are very lengthy). Taking the limit $\xi\to0$, the only term which makes a non-zero contribution has the $\xi$ raised to the exponent $4-p$, i.e., $\xi^{-4+p}$, and hence, we have 
\\
\begin{equation}
    f_{0} = \frac{3\left(1+\eta_{0}^{2}-4R\eta_{0}\eta'_{0}(R) \right)}{R},\ p = 1.
\end{equation}
\\
Since $p>0$, we proceed with step $3$ and to find Fuchs indices, we substitute the binomial 
\\
\begin{equation}
    f=\frac{3\left(1+\eta_{0}^{2}-4R\eta_{0}\eta'_{0}\right)}{R}\xi^{-1} + f_{m}\xi^{m-1}
\end{equation}
\\
into the leading terms. Upon calculations, we find four distinct values of $m$ upon solving the polynomial equation $k_{m}=0$ which is the same as that of the ODE in \ref{eq:Polynomialequation}. The Fuchs index of this PDE is $m=4$ and thus, we conclude that we are to consider four terms in the series \ref{momentumPDEPainleve}. Also, since all of the indices are non-negative, the first sufficient condition is satisfied. We proceed with step $8$ and obtain the coefficients
\\
 \begin{equation}
 \begin{array}{lr}
 -\frac{R\left(1+\eta_{0}^{2}\right)\left(1+\eta_{0} - 4R\eta_{0}\xi_{R}\right)}{8\eta_{0}}\ A_{1} = -1 + 2\eta_{0}^{2} + \eta_{0}^{4}  - R \left(1+\eta_{0}^{2}\right)\left(\eta_{0}\xi_{R} + 2R\xi_{R}^{2} - 2R\eta_{0}\xi_{RR}\right),\\\\
  
  - \frac{R\left(1+\eta_{0}^{2}\right)^{2}\left(1+\eta_{0}^{2}-4R\eta_{0}\xi_{R}\right)^{2}}{4}\ A_{2} = -2 R^{2} \left(1 + \eta_{0}^2\right)^{2} \left(-1 + 7 \eta_{0}^2\right)\xi_{R}^{2}\\\\
  + 16 R^{3} \eta_{0} (1 + \eta_{0}^2)^{2}\xi_{R}^{3} + 2 R \eta_{0}\xi_{R} \left(10 + 23 \eta_{0}^2 + 12 \eta_{0}^4 + 7 \eta_{0}^6 + 4 R^2 \eta_{0} (1 + \eta_{0}^2)^{2}\xi_{RR} \right)\\\\
  - \left(1 + \eta_{0}^{2}\right) \left(1 + 19 \eta_{0}^2 + 3 \eta_{0}^4 + \eta_{0}^6 + 4 R^2 \eta_{0} \left(2 + \eta_{0}^2 - \eta_{0}^4\right)\xi_{RR}\right), 
  \end{array}
\end{equation}
\\
and where $A_{3}$ can be arbitrary.

\subsubsection{Painlev\'e Test of PDE to higher-orders of $R$}
\noindent Calculations become very complex for the case of $\mathcal{O}(R^{2})$ and when higher-order of Reynolds number is considered in the PDE. However, for all cases the polynomial $k_{m}$ remains the one shown in \ref{eq:Polynomialequation}. Below, we list the expression of the coefficients $A_{0}$ when different orders of $R$ are considered in the PDE.
\\
\begin{equation}\label{coeffandexponent}
    \begin{array}{lr}
    \mathcal{O}(R^{2}):\ A_{0} = \frac{3\left(\left(1+\eta_{0}^{2}\right)^{2} - 4R\eta_{0}\left(1+\eta_{0}^{2}\right)\xi_{R} + 2R^{2}\left(1+3\eta_{0}^{2}\right)\xi_{R}^{2}\right)}{R\left(1+\eta_{0}^{2}\right)},\ p=1,\\\\
    
    \mathcal{O}(R^{3}):\ A_{0} = \frac{3\left(\left(1+\eta_{0}^{2}\right)^{2} - 4R\eta_{0}\left(1+\eta_{0}^{2}\right)\xi_{R} + 2R^{2}\left(1+3\eta_{0}^{2}\right)\xi_{R}^{2}\right) - 4R^{3}\eta_{0}\xi_{R}^{3}}{R\left(1+\eta_{0}^{2}\right)},\ p=1,\\\\
    
    \mathcal{O}(R^{4}):\ A_{0} = \frac{3\left(1 + \eta_{0}^{2} - 2R\eta_{0}\xi_{R} + R^{2}\xi_{R}^{2}\right)^{2}}{R\left(1+\eta_{0}^{2}\right)},\ p=1.
    \end{array}
\end{equation}
\\
Hence, the complete momentum PDE \ref{momentumfourthPDE} passes the Painlev\'e test and can be expressed as a series solution with the leading term whose coefficient and exponent are given in \ref{coeffandexponent}.

\subsection{On Route to Solutions Using the Painlev\'e Results}
\noindent We know from subsections \ref{sec:PainleveODE} and \ref{sec:PainlevePDE} that there exists a series type solution to the momentum ODE and the PDE. We first make the case for using the momentum ODE's solution prior to formulating the problem. Due to computational difficulty, we are unable to calculate the coefficients $A_{1}$, $A_{2}$ and $A_{3}$ in the series expansion of the Painlev\'e PDE solution \ref{momentumPDEPainleve} but making the assumption that $\xi_{R}=0$ simplifies the coefficient of the leading term, $A_{0}$ \ref{coeffandexponent} to that of of the momentum ODE \ref{eq:ODEcoefficient}. This turns out to be true and we will show in chapter $5$ that only if this assumption holds will the transition boundary solution resemble the Blasius solution far downstream in the large Reynolds number limit. Hence, it suffices to take the solution of the momentum ODE as given in \ref{eq:finalsolPainleveODE} for our calculations here. Taking the appropriate derivatives and under the assumption that $\xi_{R}=0$, \ref{energyeq} can be written as 
\\
\begin{equation}\label{energy with painleve stuff}
    \begin{array}{lr}
    \left(1+\eta^{2}\right)T_{\eta\eta} + 2\eta T_{\eta} + R\left(PrU - 4\eta T_{\eta} \right) + R^{2}T_{RR}\\\\
    +PrU\left(RA_{0}^{2} - 6A_{1} -2A_{0}\left(3 + RA_{2}\xi \right) + A_{2}\left(-6+RA_{2}\xi^{2}\right)  \right)\xi^{-2},
    \end{array}
\end{equation}
\\
where $\xi = \eta - \eta_{0}(R)$ and $A_{0}$, $A_{1}$, and $A_{2}$ are as given in \ref{eq:PainleveODEcoeffiients}.

\subsection{Integrability of the Energy Equation}
\noindent Consider the energy equation to order $\mathcal{O}(R)$, performing a Painlev\'e analysis substituting the leading term of the Painlev\'e series into the energy PDE and multiplying throughout by $\xi^{2+p}$, we obtain the following result
\\
\begin{equation}\label{energyPainleve}
    PrUR\xi^{2+p} + pA_{0} + p^{2}A_{0} + p\eta^{2}A_{0} + p^{2}\eta^{2}A_{0} - 2p\eta\xi A_{0} + 4pR\eta\xi A_{0}=0.
\end{equation}
\\
Now, in the limit $\xi\to0$, the value of the exponent has to be $p\geq -2$ in order to prevent a blow up and in order for the first term of \ref{energyPainleve} to give a non-zero quantity we require $p=-2$. Since $p<0$, the energy PDE to order $\mathcal{O}(R)$ fails the Painlev\'e test and the general solution is expressed as $T(\eta, R) = A_{0}\left(\eta - \eta_{0} \right)^{2}$ and has an algebraic branch point. Similarly, performing the integrability test for the energy PDE to order $\mathcal{O}(R^{2})$ as in \ref{energy with painleve stuff} we find that it too fails as the exponent $p=-2$.

\subsection{Integrability of Generalized Self-Similar Equations}
\noindent We consider a planar steady incompressible flow,i.e., 
\\
\begin{equation}\label{generalizedself-similar}
   \begin{array}{lr}
    u\frac{\partial u}{\partial x} + v\frac{\partial v}{\partial y} = U\frac{\partial U}{\partial x} + \nu\frac{\partial^{2} u}{\partial y^{2}},\\\\
    
    \frac{\partial u}{\partial x} + \frac{\partial v}{\partial y} = 0,
    \end{array}
\end{equation}
\\
where $UU_{x}=-\rho^{-1}p_{x}$ follows from Bernoulli's equation in the high Reynolds number limit. The boundary conditions are \ref{atplate} and \ref{leadingedge}.
We now introduce the stream function into the above equation and carry out a transformation from the $(x,y)$ coordinates to a dimensionless variable $\eta$ such that,
\\
\begin{equation}
\label{generalizedtransformations}
    \eta = \frac{y}{l}\frac{\sqrt{R}}{\bar{\delta}(\xi)} = \frac{\bar{y}}{\bar{\delta}(\xi)},\ \xi = \frac{x}{l},
\end{equation}
\\
where $R=Ul/\nu$ is the Reynolds number. On substituting \ref{generalizedtransformations} into \ref{generalizedself-similar}, we obtain a differential equation for the dimensionless stream function $f(\xi,\eta)$ which reads \cite{Herman}
\\
\begin{equation}\label{generalizedselfsimilarPDE}
    f''' + \alpha_{1} ff'' + \alpha_{2} - \alpha_{3} f'^{2} = \bar{\delta}^{2} \frac{U_{N}}{U}\left(f' \frac{\partial f'}{\partial \xi} - f'' \frac{\partial f}{\partial \xi}\right),
\end{equation}
\\
here $\alpha_{1}$, $\alpha_{2}$, and $\alpha_{3}$ are defined as follows
\\
\begin{equation}\label{alphaconstants}
    \alpha_{1} = \frac{\bar{\delta}}{U}\frac{d}{d\xi}\left(U_{N}\bar{\delta} \right),\ \alpha_{2} = \frac{\bar{\delta}^{2}}{U}\frac{U}{U_{N}}\frac{dU}{d\xi},\ \alpha_{3} = \frac{\bar{\delta}^{2}}{U}\frac{dU_{N}}{d\xi},
\end{equation}
\\
where $U_{N}$ is the velocity of the outer flow $U(\xi)$. The equation now reduces to an ordinary differential equation for the function $f(\eta)$, independent of $\xi$. Hence, the right-hand-side of \ref{generalizedselfsimilarPDE} vanishes and the reduced ODE, the generalized self-similar equation reads
\begin{equation}\label{generalizedselfsimilarODE}
    f''' + \alpha_{1} ff'' + \alpha_{2} - \alpha_{3} f'^{2} = 0.
\end{equation}
\\
We perform a Painlev\'e test for the generalized self-similar equation \ref{generalizedselfsimilarODE} and discuss integrability of the different types of differential equation that arise for various values of $\alpha_{1}$, $\alpha_{2}$, and $\alpha_{3}$. Substituting \ref{PainlevemomODE} in \ref{generalizedselfsimilarODE} and multiplying the resulting expression by $\xi^{3+p}$ yields
\\
\begin{equation}\label{eq:Painlevegeneralizedself-similar}
    -A_{0}p\left(2+3p+p^{2}\right) + A_{0}^{2}p\left(\alpha_{1}+p\left(\alpha_{1}+\alpha_{3}\right)\right)\xi^{1-p} + \alpha_{2}\xi^{3+p} = 0. 
\end{equation}
\\
Now, as $\xi\to 0$ for $p>0$, we notice that the only term which makes a non-zero contribution is the term multiplied by the factor $\xi^{1-p}$ and in order to prevent a blow-up we must set $p=1$. It follows from \ref{eq:Painlevegeneralizedself-similar} that the leading term of the series is given by \ref{PainlevemomODE} with,
\\
\begin{equation}
    p=1,\ A_{0} = \frac{6}{2\alpha_{1} - \alpha_{3}}.
\end{equation}
\\
Since $p>0$, the generalized self-similar equation satisfies the first sufficient condition of the integrability test. To fix the number of terms we are to consider in the series, we substitute the binomial \ref{eq:binomial} which gives the following expression for the polynomial $k_{m}$,
\\
\begin{equation}\label{generalizedself-similark_m}
    k_{m} = 6\left(2\alpha_{1} - \alpha_{3}\right) + m\left(4\alpha_{1} + \alpha_{3} \right) + 6m^{2}\left(\alpha_{3} - \alpha_{1}\right) + m^{3}\left(2\alpha_{1} - \alpha_{3} \right).
\end{equation}
\\
From the expression $k_{m} = 0$ we find the following Fuchs indices
\\
\begin{equation}\label{Fuchsgeneralizedself-similar}
    m_{1} = 1,\ m_{2,3} = \frac{8\alpha_{1} - 7\alpha_{3} \mp \sqrt{25\alpha_{3}^{2} - 16\alpha_{1}\alpha_{3} - 32\alpha_{1}^{2}}}{2\left(2\alpha_{1} - \alpha_{3}\right)}.
\end{equation}
\\
We have shown the calculations for the Fuchs indices $m_{2}$ and $m_{3}$ and have also commented on the integrability for the various types of flows in table \ref{tab:integrabilityofflows}. The immediate and interesting observation we make is that the integrability of the equation remains unaffected for any value of $\alpha_{2}$.
\\
\begin{table}
    \centering
    \begin{tabular}{c c c c c c c}
        Type & $\alpha_{1}$ & $\alpha_{3}$  & $m_{1}$ & $m_{2}$ &  $m_{3}$ & Painlev\'e Test \\ [0.25ex] 
        \hline\hline 
        Falkner-Skan & $1$ & $\beta$ & $-1$ & $\frac{8 - 7\beta - \sqrt{25\beta^{2} - 16\beta - 32}}{2\left(2 - \beta\right)}$ & $\frac{8 - 7\beta + \sqrt{25\beta^{2} - 16\beta - 32}}{2\left(2 - \beta\right)}$ & Fail $\forall
        \beta$ \\\\ 
        Reversed wedge & $-1$ & $-\beta$ & $-1$ & $\frac{-8 + 7\beta - \sqrt{25\beta^{2} - 16\beta - 32}}{2\left(-2 + \beta\right)}$ & $\frac{-8 + 7\beta + \sqrt{25\beta^{2} - 16\beta - 32}}{2\left(-2 + \beta\right)}$ &  Fail $\forall \beta$ \\\\ 
        Pohlhausen & $0$ & $1$ & $-1$ & $1$ & $6$ & Pass \\\\ 
        Blasius & $\frac{1}{2}$ & $0$ & $-1$ & $2-i\sqrt{2}$ & $2+i\sqrt{2}$ & Fail\\\\
        Free jet & $1$ & $-1$ & $-1$ & $2$ & $3$ & Pass\\\\ 
        Wall jet & $1$ & $-2$ & $-1$ & $\frac{3}{2}$ & $4$ & Pass\\\\ 
        Heimenz & $1$ & $1$ & $-1$ & $\frac{1}{2}\left(1-i\sqrt{23}\right)$ & $\frac{1}{2}\left(1+i\sqrt{23}\right)$ & Fail\\\\
        Homann  & $1$ & $\frac{1}{2}$ & $-1$ & $-\frac{1}{2}\left(3-i\sqrt{15}\right)$ & $-\frac{1}{2}\left(3+i\sqrt{15}\right)$ & Fail \\ [1ex] 
        \hline\hline
    \end{tabular}
    \caption{Integrability results of different flows. Here, the values of $\beta$ fall in the range $\beta\in \left[-0.198838,\frac{4}{3}\right]$.}
    \label{tab:integrabilityofflows}
\end{table}

\section{Lie Symmetries}
\subsection{Introduction}
\noindent We consider the momentum PDE \ref{momentumfourthPDE} where $f$ is the dependent variable and $\eta$ and $R$ are the independent variables and define the point transformation, which is a locally defined diffeomorphism, as follows \cite{Olver}
\\
\begin{equation}\label{eq:pointtransformation}
    \Gamma: (f,\eta,R)\rightarrow\left(\bar{\eta}(\eta,R,f),\bar{R}(\eta,R,f),\bar{f}(\eta,R,f) \right).
\end{equation}
\\
This point transformation maps the surface $f=\mathcal{F}(\eta,R)$ to the following surface parameterized by $\eta$ and $R$
\\
\begin{equation}
    \begin{array}{lr}
    \bar{\eta} = \bar{\eta}\left(\eta,R,\mathcal{F}(\eta,R)\right),\\\\
    \bar{R} = \bar{R}\left(\eta,R,\mathcal{F}(\eta,R)\right),\\\\
    \bar{f} = \bar{f}\left(\eta,R,\mathcal{F}(\eta,R)\right).
    \end{array}
\end{equation}
\\
We now define the Lie point transformations as follows
\\
\begin{equation}\label{eq:Liepointtransformations}
   \begin{array}{lr}
    \bar{\eta} = \eta + \epsilon \xi(\eta,R,f) + \mathcal{O}(\epsilon^{2}),\\\\ 
    \bar{R} = R + \epsilon \zeta(\eta,R,f) + \mathcal{O}(\epsilon^{2}),\\\\ 
    \bar{f} = f + \epsilon \kappa(\eta,R,f) + \mathcal{O}(\epsilon^{2}).\\\\ 
   \end{array}
\end{equation}
For these transformations, the surface $f = \mathcal{F}$ is mapped to 
\\
\begin{equation}
    \begin{array}{lr}
    \eta = \bar{\eta} - \epsilon \xi\left(\bar{\eta},\bar{R},\mathcal{F}(\bar{\eta},\bar{R})\right) + \mathcal{O}(\epsilon^{2}),\\\\
    R = \bar{R} - \epsilon \zeta\left(\bar{\eta},\bar{R},\mathcal{F}(\bar{\eta},\bar{R})\right) + \mathcal{O}(\epsilon^{2}),\\\\
    \bar{f} = \mathcal{F}(\eta,R) + \epsilon \kappa\left(\eta,R,\mathcal{F}(\eta,R)\right) + \mathcal{O}(\epsilon^{2}),
    \end{array}
\end{equation}
\\
and this corresponds to 
\\
\begin{equation}
    \bar{f} = \mathcal{F}(\eta,R) + \epsilon \mathcal{W}\left(\eta,R,\mathcal{F}(\eta,R) \right) + \mathcal{O}(\epsilon^{2}),
\end{equation}
\\
where the following is characteristic
\\
\begin{equation}
    \mathcal{W} = \kappa(\eta,R,f) - f_{\eta}\xi(\eta,R,f) - f_{R}\zeta(\eta,R,f),
\end{equation}
\\
and any other surface on which $\mathcal{W} = 0$ is considered invariant. We now define the total derivative operators that treat the dependent variable $f$ and its derivatives as functions of the independent variables $\eta$ and $R$,
\\
\begin{equation}\label{eq:Totalderivative}
    \begin{array}{lr}
    D_{\eta} = \partial_{\eta} + f_{\eta}\partial_{f} + f_{\eta\eta}\partial_{f_{\eta}} + f_{\eta R}\partial_{f_{R}} + \ldots,\\\\
    D_{R} =  \partial_{R} + f_{R}\partial_{f} + f_{RR}\partial_{f_{R}} + f_{R\eta}\partial_{f_{\eta}} + \ldots.
    \end{array}
\end{equation}
\\
We now rewrite the momentum PDE as $\mathcal{G}\left(\eta,R,[f]\right) = 0$, where $[f]$ denotes the derivatives with respect to the independent variables. The Lie transformation \ref{eq:Liepointtransformations} is a symmetry if it maps the set of solutions to itself and this gives the following symmetry condition,
\\
\begin{equation}\label{Symmetrycodition}
    \mathcal{G}\left(\bar{\eta},\bar{R},[\bar{f}]\right) = 0.
\end{equation}
\\
For this Lie symmetry, the prolonged infinitesimal generator reads
\\
\begin{equation}\label{Liegenerator}
    \mathcal{X} = \xi\partial_{\eta} + \zeta\partial_{R} + \kappa\partial_{f} + \kappa^{\eta}\partial_{f_{\eta}} + \kappa^{R}\partial_{f_{R}} +\kappa^{\eta\eta}\partial_{f_{\eta\eta}} + \kappa^{\eta R}\partial_{f_{\eta R}} + \kappa^{RR}\partial_{f_{RR}}+ \ldots,
\end{equation}
\\
where the prolongation terms read
\\
\begin{equation}
  \begin{array}{lr}
  \kappa^{\eta} = D_{\eta}\mathcal{W} + \xi f_{\eta \eta} + \zeta f_{\eta R},\\\\
  \kappa^{R} =  D_{R}\mathcal{W} + \xi f_{\eta R} + \zeta f_{R R},\\\\
  \kappa^{\eta\eta} = D_{\eta}^{2}\mathcal{W} + \xi f_{\eta \eta \eta} + \zeta f_{\eta \eta R},\\\\
  \kappa^{RR} = D_{R}^{2}\mathcal{W} + \xi f_{\eta R R} + \zeta f_{RRR},\\\\
  \kappa^{\eta R} = D_{\eta}D_{R}\mathcal{W} + \xi f_{\eta \eta R} + \zeta f_{\eta RR}, \ldots.
  \end{array}
\end{equation}
\\ 
The infinitely prolonged infinitesimal generator decomposes into $\mathcal{X} = D_{\eta} + D_{R} + \bar{\mathcal{X}}$, where the nontrivial component $\bar{\mathcal{X}}$ can be generalized as
\\
\begin{equation}
   \begin{array}{lr}
    \bar{\mathcal{X}} = \sum_{M}\left(D_{M}\mathcal{W} \right)\frac{\partial}{\partial f_{M}}\\\\
    \ \ \ \ = \mathcal{M}\partial_{f} + \left(D_{\eta}\mathcal{W}\right)\partial_{f_{\eta}} + \left((D_{R}\mathcal{W} \right)\partial_{f_{R}} + \left(D_{\eta}\mathcal{W}\right)^{2}\partial_{f_{\eta\eta}} + \left(D_{\eta}D_{R}\mathcal{W}\right)^{2}\partial_{f_{\eta R}} + \ldots,\\\\
    M = (m,n),\ D_{M} = D_{\eta}^{m}D_{R}^{n}
   \end{array}
\end{equation}
\\
The linearized Lie symmetry condition for Lie point symmetries is defined as follows
\\
\begin{equation}\label{eq:LSC}
    \mathcal{X}\left(\mathcal{G}(\eta,R,[f])\right) = 0,
\end{equation}
\\
when $\mathcal{G}(\eta,R,[f])= 0$,

\subsection{Probing for Invariant Solutions}
\noindent Our goal is to probe for an invariant solution given a PDE and its symmetries such that the solution satisfies $\mathcal{W} = 0$. Now, for Lie symmetries, the invariance condition reads
\\
\begin{equation}
    \kappa(\eta,R,f) - f_{\eta}\xi(\eta,R,f) - f_{R}\zeta(\eta,R,f) = 0,
\end{equation}
\\
which can be solved via the method of characteristics. The solution is obtained as an arbitrary locally smooth function of the first integrals of the following system (Pfaffian system)
\\
\begin{equation}\label{Pfaffian system}
    \frac{d\eta}{\xi(\eta,R,f)} = \frac{dR}{\zeta(\eta,R,f)} = \frac{df}{\kappa(\eta,R,f)}.
\end{equation}
\\
We now write down the infinitesimal Lie generator for the momentum PDE \ref{momentumfourthPDE}
\\
\begin{equation}\label{momentumPDEgenerator}
    \begin{array}{lr}
    \mathcal{X} = \xi\partial_{\eta} + \zeta\partial_{R} + \kappa\partial_{f} + \kappa^{\eta}\partial_{f_{\eta}} + \kappa^{\eta\eta}\partial_{f_{\eta\eta}} + \kappa^{\eta\eta\eta}\partial_{f_{\eta\eta\eta}} + \kappa^{\eta\eta\eta\eta}\partial_{f_{\eta\eta\eta\eta}} + \kappa^{\eta\eta R}\partial_{f_{\eta\eta R}}\\\\
    \ \ \ \ \ + \kappa^{\eta\eta\eta R}\partial_{f_{\eta\eta\eta R}} + \kappa^{R}\partial_{f_{R}} + \kappa^{\eta\eta RR}\partial_{f_{\eta\eta RR}} + \kappa^{\eta RR}\partial_{f_{\eta RR}} + \kappa^{RR}\partial_{f_{RR}} + \kappa^{RRR}\partial_{f_{RRR}}\\\\
    \ \ \ \ \ + \kappa^{\eta RRR}\partial_{f_{\eta RRR}} + \kappa^{RRRR}\partial_{f_{RRRR}}.
    \end{array}
\end{equation}
\\
For the sake of simplicity, we first operate on the momentum PDE to order $\mathcal{O}(R)$ with the Lie generator. Doing the same yields the following 
\\
\begin{equation}\label{Lieequation}
    \begin{array}{lr}
    \xi\left(4\eta \left(1+\eta^{2}\right)f_{\eta\eta\eta\eta} + 8\left(1+3\eta^{2} \right)f_{\eta\eta\eta} + 24\eta f_{\eta\eta}\right)\\\\
    + R\xi\left(2ff_{\eta\eta} + 2\eta\left(ff_{\eta\eta\eta} + f_{\eta}f_{\eta\eta} \right)  -24\eta f_{\eta\eta R} - 4f_{\eta\eta\eta R}\right)\\\\
    + \zeta\left(2\eta ff_{\eta\eta} + \left(1+\eta^{2}\right)\left(ff_{\eta\eta\eta} + f_{\eta}f_{\eta\eta} \right)  -4\left(1+3\eta^{2}\right)f_{\eta\eta R} - 4\eta\left(1+\eta^{2}\right)f_{\eta\eta\eta R} \right)\\\\
    + \kappa R\left(1+\eta^{2}\right)f_{\eta\eta\eta} + \kappa^{\eta\eta}\left(4\left(1+3\eta^{2}\right) +R\left(1+\eta^{2}\right)f_{\eta\eta}\right) + \kappa^{\eta\eta\eta}\left(8\eta\left(1+\eta^{2}\right)+R\left(1+\eta^{2}\right)f\right) \\\\
    + \kappa^{\eta\eta\eta\eta}\left(1+\eta^{2}\right)^{2}
    - 4\kappa^{\eta\eta R}\left(1+3\eta^{2}\right)R - 4\kappa^{\eta\eta\eta R}\eta \left(1+\eta^{2}\right)R = 0,
    \end{array}
\end{equation}
\\
where the prolongation terms read
\begin{equation}\label{prolongation}
    \begin{array}{lr}
    \kappa^{\eta} = \kappa_{\eta} + \left(\kappa_{f} - \xi_{\eta} \right)f_{\eta} -\xi_{f}f_{\eta}^{2} -\zeta_{\eta}f_{R} - \zeta_{f}f_{\eta}f_{R},\\\\
    
    \kappa^{R} = \kappa_{R} +\left(\kappa_{f} - \zeta_{\eta}\right)\xi_{R} - \zeta_{f}f_{R}^{2} - \xi_{R}f_{\eta} - \xi_{f}f_{R}f_{\eta},\\\\
    
    \kappa^{\eta\eta} = \kappa_{\eta\eta} +\left(2\kappa_{\eta f} -\xi_{\eta\eta}\right)f_{\eta} - \zeta_{\eta\eta}f_{R} + \left(\kappa_{ff} - 2\xi_{\eta f} \right)f_{\eta}^{2} - 2\zeta_{\eta f}f_{\eta}f_{R} - \xi_{ff}f_{\eta}^{3} - \zeta_{ff}f_{\eta}^{2}f_{R}\\\\
    \ \ \ \ \ \ + \left(\kappa_{f} - 2\xi_{\eta}\right)f_{\eta\eta} - 2\zeta_{\eta}f_{\eta R} - 3\xi_{f}f_{\eta}f_{\eta\eta} - \zeta_{f}f_{R}f_{\eta\eta} - 2\zeta_{f}f_{\eta}f_{\eta R},\\\\
    
    \kappa^{\eta\eta\eta} = \kappa_{\eta\eta\eta} - \zeta_{f}f_{R}f_{\eta\eta\eta} - 3\zeta_{\eta}f_{\eta\eta R} - 3\zeta_{\eta f}f_{\eta\eta}f_{R} - \zeta_{\eta\eta\eta}f_{R} + \kappa_{f}f_{\eta\eta\eta} + 3\kappa_{\eta f}f_{\eta\eta}\\\\
    \ \ \ \ \ \ - 3f_{\eta R}\left(\zeta_{f}f_{\eta\eta} +  \zeta_{ff}f_{\eta}^{2} + 2\zeta_{\eta f}f_{\eta} + \zeta_{\eta\eta} \right) - 3\xi_{f}f_{\eta\eta}^{2} - \xi_{fff}f_{\eta}^{4} - 3\xi_{\eta}f_{\eta\eta\eta}\\\\
    \ \ \ \ \ \ + f_{\eta}^{3}\left(\kappa_{fff} -3\eta_{\eta ff}-\zeta_{fff}f_{R} \right) - 3\xi_{\eta\eta}f_{\eta\eta} - 3f_{\eta}^{2}\left(\zeta_{\eta ff}f_{R}- \kappa_{\eta ff} + 2\xi_{ff}f_{\eta\eta} + \xi_{\eta\eta f}\right)\\\\
    \ \ \ \ \ \ - f_{\eta}\left(3\zeta_{f}f_{\eta\eta R} + 3f_{r}\left(\zeta_{ff}f_{\eta\eta} + \zeta_{\eta\eta f} \right) - 3\kappa_{ff}f_{\eta\eta} -3 \kappa_{\eta\eta f} + 4\xi_{f}f_{\eta\eta\eta} + 9\xi_{\eta f}f_{\eta\eta} + \xi_{\eta\eta\eta}\right),\\\\
    
    \kappa^{\eta\eta\eta\eta} = \kappa_{\eta\eta\eta\eta} - \zeta_{f}\left(4f_{\eta}f_{\eta\eta\eta R} + f_{R}f_{\eta\eta\eta\eta}\right) - 2\zeta_{ff}\left(3f_{\eta}^{2}f_{\eta\eta R} + 2f_{\eta}f_{\eta\eta\eta} \right) - \zeta_{ffff}f_{R}f_{\eta}^{4}\\\\
    \ \ \ \ \ \ - 4\zeta_{\eta}f_{\eta\eta\eta R} - 12\zeta_{\eta f}f_{\eta}f_{\eta\eta R} - 4\zeta_{\eta f}f_{R}f_{\eta\eta\eta} - 4\zeta_{\eta fff}f_{R}f_{\eta}^{3} - 6\zeta_{\eta\eta}f_{\eta\eta R} - 6\zeta_{\eta\eta ff}f_{R}f_{\eta}^{2}\\\\
    \ \ \ \ \ \ -4f_{\eta R}\left(\zeta_{f}f_{\eta\eta\eta} + \zeta_{fff}f_{\eta}^{3} + 3\zeta_{\eta ff}f_{\eta}^{2} + 3\zeta_{\eta\eta f}f_{\eta} + \zeta_{\eta\eta\eta}\right) - 4\zeta_{\eta\eta\eta f}f_{R}f_{\eta} - \zeta_{\eta\eta\eta\eta}f_{R} + \kappa_{f}f_{\eta\eta\eta\eta}\\\\
    \ \ \ \ \ \ + 4\kappa_{ff}f_{\eta}f_{\eta\eta\eta} + \kappa_{ffff}f_{\eta}^{4} + 4\kappa_{\eta f}f_{\eta\eta\eta} + 4\kappa_{\eta fff}f_{\eta}^{3} + 6\kappa_{\eta\eta ff}f_{\eta}^{2} + 4\kappa_{\eta\eta\eta f}f_{\eta} - 5\xi_{f}f_{\eta}f_{\eta\eta\eta\eta}\\\\
    \ \ \ \ \ \ -10\xi_{ff}f_{\eta}^{2}f_{\eta\eta\eta} - \xi_{ffff}f_{\eta}^{5} - 4\xi_{\eta}f_{\eta\eta\eta\eta} - 16 \xi_{\eta f}f_{\eta}f_{\eta\eta\eta}\\\\
    \ \ \ \ \ \ -3f_{\eta\eta}^{2}\left(\zeta_{ff}f_{R} -\kappa_{ff} + 5\xi_{ff}f_{\eta} + 4\xi_{\eta f}\right) - 4\xi_{\eta fff}f_{\eta}^{4} - 6\xi_{\eta\eta}f_{\eta\eta\eta} - 6\xi_{\eta\eta ff}f_{\eta}^{3}\\\\
    \ \ \ \ \ \ -2f_{\eta\eta}\left(3\zeta_{f}f_{\eta\eta R} + 6\zeta_{\eta f}f_{\eta R} + 3\zeta_{\eta\eta f}f_{R} - 3\kappa_{\eta\eta f} + 5\xi_{f}f_{\eta\eta\eta} + 5\xi_{fff}f_{\eta}^{3} \right)\\\\
    \ \ \ \ \ \ -2f_{\eta\eta}\left(3f_{\eta}^{2}\left(\zeta_{fff}f_{R} - \kappa_{fff} + 4\xi_{\eta ff} \right) + 2\xi_{\eta\eta\eta}\right)\\\\
    \ \ \ \ \ \ -2f_{\eta\eta}f_{\eta}\left(6\zeta_{ff}f_{\eta R} + 6\zeta_{\eta\ ff}f_{R} - 6\kappa_{\eta ff} + 9\xi_{\eta\eta f} \right) - 4\xi_{\eta\eta\eta f}f_{\eta}^{2} - \xi_{\eta\eta\eta\eta}f_{\eta},
    \end{array}
\end{equation}
\\
and the mixed prolongation terms read
\\
\begin{equation}\label{mixedprolongation}
    \begin{array}{lr}
    \kappa^{\eta\eta R} =\kappa_{\eta\eta R} -\zeta_{f}\left(2f_{\eta R}^2 + f_{RR}f_{\eta\eta} + 2f_{R}f_{\eta\eta R}\right) + \zeta_{ff}f_{R}^{2}f_{\eta\eta} - \zeta_{R}f_{\eta\eta R} - \zeta_{R f}f_{R}f_{\eta\eta} - 2\zeta_{\eta}f_{\eta RR}\\\\
    \ \ \ \ \  - \zeta_{\eta\eta}f_{RR} - \zeta_{\eta\eta f}f_{R}^{2} - \zeta_{\eta\eta R}f_{R} + \kappa_{f}f_{\eta\eta R} + \kappa_{ff}f_{R}f_{\eta\eta} + \kappa_{Rf}f_{\eta\eta} + \kappa_{\eta\eta f}f_{R}\\\\
    \ \ \ \ \ \ - \xi_{f}f_{R}f_{\eta\eta\eta} - \xi_{R}f_{\eta\eta\eta} - f_{\eta}^{3}\left(\xi_{fff}f_{R} + \xi_{Rff} \right) - 2\xi_{\eta}f_{\eta\eta R} - 2\xi_{\eta f}f_{R}f_{\eta\eta} - 2\xi_{\eta R}f_{\eta\eta}\\\\
    \ \ \ \ \ \ - f_{\eta}^{2}\left(\zeta_{ff}f_{RR} + \zeta_{fff}f_{R}^{2} - \kappa_{Rff} + f_{R}\left(\zeta_{Rff} - \kappa_{fff} + 2\xi_{\eta ff} \right) +2 \xi_{\eta Rf}\right)\\\\
    \ \ \ \ \ \ - f_{\eta R}\left(4f_{R}\left(\zeta_{ff}f_{\eta} + \zeta_{\eta f}\right) + 2\zeta_{\eta R} - 2\kappa_{\eta f} + 3\xi_{f}f_{\eta\eta} + 3\xi_{ff}f_{\eta}^{2} + 2f_{\eta}\left(\zeta_{Rf} - \kappa_{ff} + 2\xi_{\eta f}) +\xi_{\eta\eta}\right) \right)\\\\
    \ \ \ \ \ \ -f_{\eta}\left(2\zeta_{f}f_{\eta RR} + 2\zeta_{\eta f} f_{RR} + 2\zeta_{\eta ff}f_{R}^{2} + 2\zeta_{\eta Rf}f_{R} - 2\kappa_{\eta ff}f_{R} - 2\kappa_{\eta Rf}\right)\\\\
    \ \ \ \ \ \ \ - f_{\eta}\left(3\xi_{f}f_{\eta\eta R} + 3\xi_{ff}f_{R}f_{\eta\eta} +3 \xi_{Rf}f_{\eta\eta} + \xi_{\eta\eta f}f_{R} + \xi_{\eta\eta R} \right),\\\\
    
    \kappa^{\eta\eta\eta R} = \kappa_{\eta\eta\eta R}  - \zeta_{f}\left( 3f_{\eta}f_{\eta\eta RR} + f_{RR}f_{\eta\eta\eta} - 2f_{R}f_{\eta\eta\eta R}\right) - \zeta_{ff}\left(3f_{RR}f_{\eta}f_{\eta\eta} - 6f_{R}f_{\eta}f_{\eta\eta R} - f_{R}^{2}f_{\eta\eta\eta}\right)\\\\
    \ \ \ \ \ \ \ - \zeta_{fff}\left(f_{RR}f_{\eta}^{3} - 3f_{R}^{2}f_{\eta}f_{\eta\eta}\right) - \zeta_{ffff}f_{R}^{2}f_{\eta}^{3} - \zeta_{R}f_{\eta\eta\eta R} - \zeta_{Rf}\left(3f_{\eta}f_{\eta\eta R} - f_{R}f_{\eta\eta\eta}\right)\\\\
    \ \ \ \ \ \ \ - 3\zeta_{Rff}f_{R}f_{\eta}f_{\eta\eta} - \zeta_{Rfff}f_{R}f_{\eta}^{3} - 3\zeta_{\eta}f_{\eta\eta RR} - 3\zeta_{\eta f}\left(f_{RR}f_{\eta\eta} + 2f_{R}f_{\eta\eta R}\right) - 6f_{\eta R}^{2}\left(\zeta_{ff}f_{\eta} + \zeta_{\eta f}\right)\\\\
    \ \ \ \ \ \ \ - 3\zeta_{\eta ff}\left(f_{RR}f_{\eta}^{2} + f_{\eta\eta}f_{R}^{2}\right) - 3\zeta_{\eta fff}f_{R}^{2}f_{\eta}^{2}, - 3\zeta_{\eta R}f_{\eta\eta R} - 3\zeta_{\eta Rf}f_{R}f_{\eta\eta} - 3\zeta_{\eta Rff}f_{R}f_{\eta}^{2}\\\\
    \ \ \ \ \ \ \ -3f_{\eta RR}\left(\zeta_{f}f_{\eta\eta} + \zeta_{ff}f_{\eta}^{2} + 2\zeta_{\eta f}f_{\eta} + \zeta_{\eta\eta}\right) - 3\zeta_{\eta\eta f}f_{RR}f_{\eta} - 3\zeta_{\eta\eta ff}f_{R}^{2}f_{\eta} - 3\zeta_{\eta\eta Rf}f_{R}f_{\eta}\\\\
    \ \ \ \ \ \ \ - \zeta_{\eta\eta\eta}f_{RR} - \zeta_{\eta\eta\eta f}f_{R}^{2} - \zeta_{\eta\eta\eta R}f_{R} + \kappa_{f}f_{\eta\eta\eta R} + \kappa_{ff}\left(3f_{\eta}f_{\eta\eta R} + f_{R}f_{\eta\eta\eta} \right) + 3\kappa_{fff}f_{R}f_{\eta}f_{\eta\eta}\\\\
    \ \ \ \ \ \ \ + \kappa_{ffff}f_{R}f_{\eta}^{3} + \kappa_{Rf}f_{\eta\eta\eta} + 3\kappa_{Rff}f_{\eta}f_{\eta\eta} + \kappa_{Rfff}f_{\eta}^{3} + 3\kappa_{\eta f}f_{\eta\eta R} +3 \kappa_{\eta ff}f_{R}f_{\eta\eta} + 3\kappa_{\eta fff}f_{R}f_{\eta}^{2}\\\\ 
    \ \ \ \ \ \ \ + 3\kappa_{\eta Rf}f_{\eta\eta} + 3\kappa_{\eta Rff}f_{\eta}^{2} + 3\kappa_{\eta\eta ff}f_{R}f_{\eta} + 3\kappa_{\eta\eta Rf}f_{\eta} + \kappa_{\eta\eta\eta f}f_{R}\\\\
    \ \ \ \ \ \ \ - \xi_{f}\left(6f_{\eta\eta}f_{\eta\eta R} + 4f_{\eta}f_{\eta\eta\eta R} + f_{R}f_{\eta\eta\eta\eta} \right) - \xi_{ff}\left(3f_{R}f_{\eta\eta}^{2} + 6f_{\eta}^{2}f_{\eta\eta R} + 4f_{R}f_{\eta}f_{\eta\eta\eta} \right)\\\\
    \ \ \ \ \ \ \ + 6\xi_{fff}f_{R}f_{\eta}^{2}f_{\eta\eta} - \xi_{ffff}f_{R}f_{\eta}^{4} - \xi_{R}f_{\eta\eta\eta\eta} - \xi_{Rf}\left(3f_{\eta\eta}^{2} + 4f_{\eta}f_{\eta\eta\eta}\right) - 6\xi_{Rff}f_{\eta}^{2}f_{\eta\eta} - \xi_{Rfff}f_{\eta}^{4}\\\\
    \ \ \ \ \ \ \ - 3\xi_{\eta}f_{\eta\eta\eta R} - \xi_{\eta f}\left(9f_{\eta}f_{\eta\eta R} + 3f_{R}f_{\eta\eta\eta} \right) - 9\xi_{\eta ff}f_{R}f_{\eta}f_{\eta\eta} - 3\xi_{\eta fff}f_{R}f_{\eta}^{3} - 3\xi_{\eta R}f_{\eta\eta\eta}\\\\
    \ \ \ \ \ \ \ - 9\xi_{\eta Rf}f_{\eta}f_{\eta\eta} - 3\xi_{\eta Rff}f_{\eta}^{3} - 3\xi_{\eta\eta}f_{\eta\eta R} - 3\xi_{\eta\eta f}f_{R}f_{\eta\eta} - 3\xi_{\eta\eta ff}f_{R}f_{\eta}^{2} - 3\xi_{\eta\eta R}f_{\eta\eta} - 3\xi_{\eta\eta Rf}f_{\eta}^{2}\\\\
    \ \ \ \ \ \ \ -f_{\eta R}\left(6\zeta_{f}f_{\eta\eta R} + 3\zeta_{Rf}f_{\eta\eta} + 3\zeta_{Rff}f_{\eta}^{2} + 6\zeta_{\eta Rf}f_{\eta} + 6f_{R}\left(\zeta_{ff}f_{\eta\eta} + \zeta_{fff}f_{\eta}^{2} + 2\zeta_{\eta ff}f_{\eta} + \zeta_{\eta\eta f} \right)\right)\\\\
    \ \ \ \ \ \ \ -f_{\eta R}\left(3\zeta_{\eta\eta R} - 3\kappa_{ff}f_{\eta\eta} - \kappa_{fff}f_{\eta}^{2} - 6\zeta_{\eta ff}f_{\eta} - 3\kappa_{\eta\eta f} + 4\xi_{f}f_{\eta\eta\eta} + 12\xi_{ff}f_{\eta}f_{\eta\eta} + 4\xi_{fff}f_{\eta}^{3} \right)\\\\
    \ \ \ \ \ \ \ -f_{\eta R}\left(9f_{\eta\eta}\xi_{\eta f} + 9\xi_{\eta ff}f_{\eta}^{2} + 6\xi_{\eta\eta f}f_{\eta} + \xi_{\eta\eta\eta} \right)- \xi_{\eta\eta\eta f}f_{R}f_{\eta} - \xi_{\eta\eta\eta R}f_{\eta}.
    \end{array}
\end{equation}
\\
When the terms $f_{\eta\eta\eta\eta}$ and $f_{\eta\eta\eta R}$ are replaced by \ref{Lieequation} in the momentum PDE to order $\mathcal{O}(R)$, the highest-order derivative terms in the Lie symmetry condition \ref{Lieequation} has factors $f_{\eta\eta\eta R}$ which when written yields
\\ 
\begin{equation}\label{factor1}
    \left(1+\eta^{2}\right)^{2}\left(- 8\zeta_{\eta} \right) - 4R\eta\left(1+\eta^{2} \right)\left(\zeta_{R} + \kappa_{f} - 7\xi_{\eta}\right) -4R\xi - 4\eta \left(1+\eta^{2}\right)\zeta= 0,
\end{equation}
\\
the factor of $f_{\eta\eta\eta\eta}$ which when written yields 
\\
\begin{equation}\label{factor2}
    \left(1+\eta^{2}\right)^{2}\left(-\zeta_{R} + \kappa_{f} - 9\xi_{\eta}\right) - 4R\left(1+\eta^{2} \right)\eta\left(-2 \xi_{R}\right) + 4\eta\left(1+\eta^{2} \right)\xi  = 0,
\end{equation}
\\
the factor of $f_{\eta\eta RR}$ which when written yields
\\
\begin{equation}\label{factor3}
   \begin{array}{lr}
   - 4R\left(1+\eta^{2}\right)\eta\left(-6\zeta_{\eta}\right)= 0,\\\\
   \zeta_{\eta} = 0.
   \end{array}
\end{equation}
\\
This result implies that $\zeta = C_{1}=$constant and removes many terms from the Lie symmetry condition equation and when simplified, we obtain 
\\
\begin{equation}\label{kappaexpression}
    \kappa_{f} = \xi_{\eta}
\end{equation}
\\
We now use \ref{kappaexpression} and \ref{factor3} in \ref{factor2} to get
\\
\begin{equation}\label{reducedfactor2}
    -10\left(1+\eta^{2}\right)\xi_{\eta} + 8R\eta \xi_{R} + 4\eta \xi = 0,
\end{equation}
\\
which is a linear partial differential equation for $\xi$ in $\eta$ and $R$. Equation \ref{reducedfactor2} can be solved via the method of characteristics to give
\\
\begin{equation}\label{xi solution}
    \xi = \left(1+\eta^{2} \right)^{1/5}k\left(C_{2}\right),\ C_{2} = R\left(1+\eta^{2}\right)^{2/5}.
\end{equation}
\\
Using the result of \ref{xi solution} in \ref{kappaexpression}, we obtain 
\\
\begin{equation}\label{kappa solution}
  \kappa = \frac{2\eta}{5\left(1+\eta^{2}\right)^{4/5}}\left(k\left(C_{2}\right) + 2R\left(1+\eta^{2}\right)^{2/5}k'\left(C_{2}\right)\right)f(\eta, R).  
\end{equation}
\\
Using the \ref{kappa solution}, \ref{xi solution} and \ref{factor3} in the Pfaffian system \ref{Pfaffian system}, we can obtain the solution to $f(\eta, R)$ in terms of an integral
\\
\begin{equation}
    ln\left(f(\eta, R) \right) = \frac{2\eta}{5\left(1+\eta^{2}\right)^{4/5}} \int\left(k\left(C_{2}\right) + 2R\left(1+\eta^{2}\right)^{2/5}f'\left(C_{2}\right)\right)\ dR.
\end{equation}
\\
To further simplify the problem, let us assume that the initial condition $\xi=C_{3}(\eta)$ holds at $R=\alpha$. Then, \ref{xi solution} reads $C_{3}(\eta) = \left(1+\eta^{2} \right)^{1/5}k\left(C_{2}\right)$, with $C_{2} = \alpha\left(1+\eta^{2}\right)^{2/5}$. This expression can now be inverted and expressed via a dummy variable, say $r$, as follows
\\
\begin{equation}
    k(r) = \left(1+r^{2}\right)^{1/5}C_{3}\left(\alpha \left(1+r^{2}\right)^{2/5} \right),
\end{equation}
\\
and substituting $r=C_{2}$, we have 
\\
\begin{equation}
  \xi = \left(1+\eta^{2} \right)^{1/5}\left(1+R^{2}\left(1+\eta^{2}\right)^{4/5} \right)^{1/5} g\left(C_{4} \right),\ C_{4} = C_{3}\left(\alpha\left(1+R^{2}\left(1+\eta^{2}\right)^{4/5}\right)^{2/5} \right).  
\end{equation}
\\
This form can now be used to find the expression for $f(\eta, R)$ after fixing the form of $g\left(C_{4}\right)$. 
\newline
\newline
\noindent We now find all the prolongation terms in \ref{momentumPDEgenerator} when the same acts on the momentum PDE to all orders of Reynolds number. The explicit expressions of the prolongation terms are given in the appendix (under section \ref{appendix2}). Below is the expression obtained when the Lie generator acts on the momentum PDE to all orders of $R$,
\\
\begin{equation}
    \begin{array}{lr}
     \xi\left(4\eta \left(1+\eta^{2}\right)f_{\eta\eta\eta\eta} + 8\left(1+3\eta^{2} \right)f_{\eta\eta\eta} + 24\eta f_{\eta\eta}\right)\\\\
    + R\xi\left(2ff_{\eta\eta} + 2\eta\left(ff_{\eta\eta\eta} + f_{\eta}f_{\eta\eta} \right)  -24\eta f_{\eta\eta R} - 4f_{\eta\eta\eta R}\right)\\\\
    + \zeta\left(2\eta ff_{\eta\eta} + \left(1+\eta^{2}\right)\left(ff_{\eta\eta\eta} + f_{\eta}f_{\eta\eta} \right)  -4\left(1+3\eta^{2}\right)f_{\eta\eta R} - 4\eta\left(1+\eta^{2}\right)f_{\eta\eta\eta R} \right)\\\\
    + \kappa R\left(1+\eta^{2}\right)f_{\eta\eta\eta} + \kappa^{\eta\eta}\left(4\left(1+3\eta^{2}\right) +R\left(1+\eta^{2}\right)f_{\eta\eta}\right) + \kappa^{\eta\eta\eta}\left(8\eta\left(1+\eta^{2}\right)+R\left(1+\eta^{2}\right)f\right) \\\\
    + \kappa^{\eta\eta\eta\eta}\left(1+\eta^{2}\right)^{2}
    - 4\kappa^{\eta\eta R}\left(1+3\eta^{2}\right)R - 4\kappa^{\eta\eta\eta R}\eta \left(1+\eta^{2}\right)R\\\\
    \color{blue}{+\xi\left(2\left(f_{R}f_{\eta \eta} - ff_{\eta\eta R} \right) - 2\eta\left(f_{\eta}f_{\eta\eta R} - f_{R}f_{\eta\eta\eta} \right) +12\eta f_{\eta\eta RR}\right)R^{2}}\\\\
    \color{blue}{+2R\zeta\left(2\eta \left(f_{R}f_{\eta \eta} - ff_{\eta\eta R}\right) - \left(1+\eta^{2}\right)\left(f_{\eta}f_{\eta\eta R} - f_{R}f_{\eta\eta\eta} \right) + 2\left(1+3\eta^{2}\right)f_{\eta\eta RR}\right)}\\\\
    \color{blue}{-2\kappa R^{2}\eta f_{\eta\eta R} +\kappa^{\eta\eta}R^{2}\eta f_{R} + \kappa^{\eta\eta\eta}R^{2}\left(1+\eta^{2}\right)f_{R} - \kappa^{\eta}\left(1+\eta^{2}\right)R^{2}f_{\eta\eta R} + 2\kappa^{\eta\eta RR}R^{2}\left(1+3\eta^{2}\right)}\\\\
    \color{blue}{+\kappa^{R}R^{2}\left(2\eta f_{\eta\eta} - \left(1+\eta^{2}\right)f_{\eta\eta\eta}\right) - \kappa^{\eta\eta\eta}R^{2}\left(2\eta f + \left(1+\eta^{2}\right)f_{\eta} \right)}\\\\
    \color{orange}{+\xi\left(2\left(f_{\eta}f_{\eta RR} - f_{R}f_{\eta \eta R} \right) - 4f_{\eta RRR}\right)+\kappa^{\eta}R^{3}\left(2\eta f_{\eta RR} - 3f_{RR} \right) + \kappa R^{3}f_{\eta RR} - 2\kappa^{R}R^{3}\eta f_{\eta\eta R}}\\\\
    \color{orange}{+ 3R^{2}\zeta\left(2\eta\left(f_{\eta}f_{\eta RR} - f_{R}f_{\eta \eta R} \right) + ff_{\eta RR} - 3f_{\eta}f_{RR}+ 4f_{RRR} - 4\eta f_{\eta RRR} \right)}\\\\ \color{orange}{+ \kappa^{\eta RR}R^{3}\left(2\eta f_{\eta} + f \right) - 2\kappa^{\eta \eta R}R^{3}\eta f_{R} - 3\kappa^{RR}R^{3}f_{\eta} + 4\kappa^{RRR}R^{3} - 4\kappa^{\eta RRR}R^{3}\eta}\\\\
    \color{purple}{+4\zeta R^{3}\left(f_{RRRR} + f_{R}f_{\eta RR} - f_{\eta}f_{RRR} \right) + \kappa^{R}R^{4}f_{\eta RR} - \kappa^{\eta}R^{4}f_{RRR} - \kappa^{RRR}R^{4}f_{\eta}}\\\\
    \color{purple}{+ \kappa^{\eta RR}R^{4}f_{R} + \kappa^{RRRR}R^{4}} = 0,
    \end{array}
\end{equation}
\\
where the terms coloured blue represent the result of the Lie generator action on terms of order $\mathcal{O}\left(R^{2}\right)$, while the terms coloured orange represent the result on the terms of order  $\mathcal{O}\left(R^{3}\right)$, and the terms coloured purple represent the result on the terms of order $\mathcal{O}\left(R^{4}\right)$. The highest-order derivative terms in the Lie symmetry condition have common factors $f_{\eta\eta\eta R}$, $f_{\eta\eta\eta\eta}$, $f_{\eta\eta RR}$, $f_{\eta RRR}$, and $f_{RRRR}$ which when written respectively yield the following equations
\\
\begin{equation}\label{completeLielist}
   \begin{array}{lr}
    -4\zeta \eta \left(1+\eta^{2} \right) - 4R\xi +  \left(1+\eta^{2} \right)^{2}\left(-8\zeta_{\eta}\right) + 2R^{2} \left(1+3\eta^{2} \right)\left(\xi - 4\xi_{R}\right)\\\\
    - 4R\eta  \left(1+\eta^{2} \right)\left(-3\zeta_{R} +\kappa_{f} - 7\xi_{\eta}\right) = 0, \\\\
    
    4R\zeta \left(1+3\eta^{2} \right) + 12R^{2}\eta \xi + 24R\eta \left(1+\eta^{2} \right)\zeta_{\eta} + 2R^{2}\left(1+3\eta^{2} \right)\left(\zeta - 5\zeta_{R} + \kappa_{f} - 5\xi_{\eta} \right) = 0, \\\\
    
    -8R^{2}\left(1+3\eta^{2} \right)\zeta_{\eta} - 8R^{4}\xi_{R} - 4\left(R^{3}\eta + 3R^{2}\zeta \eta + \xi + R^{2}\eta\left(-7\zeta_{R} + \kappa_{f} - 3\xi_{\eta} \right) \right)=0, \\\\
    
    4\eta \xi + 8R\eta\xi_{R} + \left(1+\eta^{2} \right)\left(-\zeta_{R} + \kappa_{f} - 3\xi_{\eta} \right) = 0,\\\\
    
    4R^{3}\zeta -4R^{3}\eta + R^{4} = 0.
    \end{array}
\end{equation}
\\
From the last equation we can fix the value of $\zeta$ and its derivatives as 
\\
\begin{equation}\label{zetaall Lie}
    \zeta = \frac{1}{4}\left(4\eta - R\right),\ \zeta_{R} =- \frac{1}{4},\ \zeta_{\eta} = 1.
\end{equation}
\\
We can now, from \ref{completeLielist}, substitute for $\xi_{R}$ from the fourth equation into the third and similar repeat the same for equation two and one. Then we can substitute for $\xi_{\eta}$ from any one of the newly obtained equations and finally use the result of \ref{zetaall Lie} to obtain a relation between $\kappa_{f}$ and $\xi$. Using this relation in the fourth equation of \ref{completeLielist} helps us obtain a PDE for $\xi$ in variables $\eta$ and $R$.  

\section{The Boundary-Matching Approach}
\begin{figure}
    \centering
    \includegraphics[width=140mm]{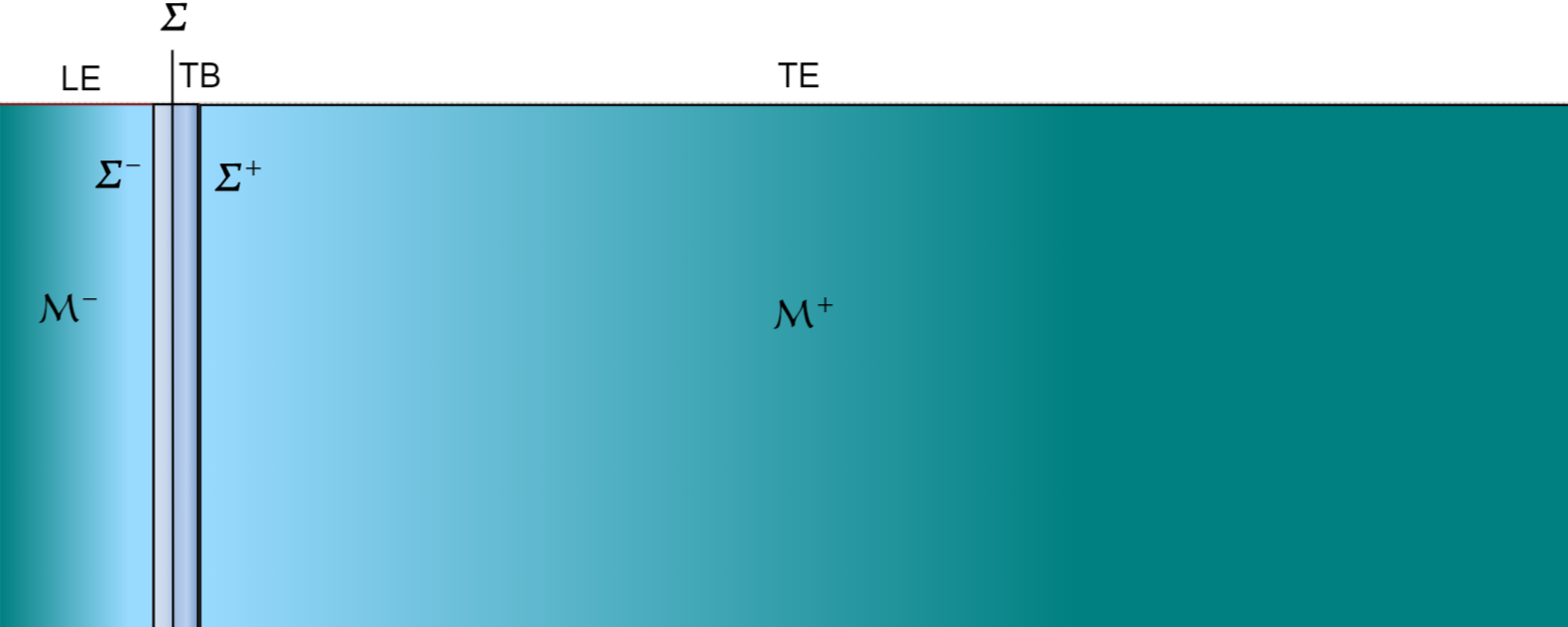}
    \caption{This figure shows the partitioning of the flat plate into the leading edge domain (LE) $\mathcal{M}^{-}$, the inner transition boundary $\Sigma^{-}$, the transition boundary (TB) $\Sigma$, the outer transition boundary $\Sigma^{+}$, and the trailing edge domain (TE) $\mathcal{M}^{+}$.}
    \label{fig:LETBTE}
\end{figure}

\noindent Let $\mathcal{M}$ denote the entire set of solutions on the flat plate among which $\mathcal{M}^{-}\subset \mathcal{M}$ is the set of solutions at the leading edge\footnote{Note that $\mathcal{M}^{+}$ is the set space where the Blasius equation and it's solution holds since this deals with the trailing edge of the plate} (see figure \ref{fig:LETBTE}). Let $\Sigma$ denote the boundary separating $\mathcal{M}^{-}$ and $\mathcal{M}^{+}$ and let the size of the leading edge subset be measured by a bookkeeping parameter $\epsilon$. 
\\
\\
Note that the solutions in $\mathcal{M}^{-}$ strictly obey the condition that $\eta\leq\epsilon$. Since the solution also depends upon Reynolds number, the parameter $\epsilon$ will be a function of $R$, i.e., $\epsilon = g(R)$, this is so as to not keep its value arbitrary. Our aim is now to develop a transition solution at the boundary of the leading edge and the trailing edge and fix constraints by matching across the boundary. We know from performing the Painlev\'e test to the momentum PDE that the solution takes the form as mentioned in \ref{momentumPDEPainleve}. We now consider the trailing edge solution to be given in terms of the following power series as was given by Blasius \cite{Blasius}
\\
\begin{equation}\label{eq:Blasiuspowerseriessolution}
    \begin{array}{lr}
    f(\eta) = \sum_{n=0}^{\infty}\left(-\frac{1}{2}\right)^{n}\frac{A_{n}\sigma^{n+1}}{\left(3n + 2\right)!}\eta^{3n+2},\\\\\\
    A_{n} = \left\{
\begin{array}{c l}	
     1 & n=0\ \&\ n=1\\\\
     \sum_{r=0}^{n-1}\left(\begin{array}{cc}
        3n-1  &  \\
         3r & 
     \end{array} \right)A_{r}A_{n-r} & n\geq 2
\end{array}\right.
    \end{array}
\end{equation}
\\
where $\sigma = f''(0)$. The rough outline of the procedure is as follows, we match the leading edge solution and the Blasius solution to the boundary solution ansatz which we assume to be of the form as
\\\begin{equation}\label{eq:Ansatz}
    f(\eta,R) = \sum_{m,n=0}^{\infty}\alpha_{n}(R)\ \beta_{m}(\eta),
\end{equation}
\\
\noindent where
\begin{equation}\label{eq:Ansatz's coefficients}
     \alpha_{n}(R) = \sum_{j=0}^{\infty}\zeta_{n,j}\ R^{j},\ \beta_{m}(\eta) = \frac{1}{\eta^{p}}\sum_{k=0}^{\infty}\lambda_{m,k}\ \eta^{k},
\end{equation}
\\
with $p=1$ (simple pole-type singularity). We rewrite all the solutions across the junction by taking the right-hand and left-hand limits of $\eta$ respectively, i.e.,
\\
\begin{equation}
\begin{array}{lr}
    f^{LE}(\eta, R) \equiv \lim^{\mathcal{M}^{-}}_{\eta \to \epsilon(\Sigma)} \xi^{-p}\sum_{n=0}^{\infty}f_{n}(R)\xi^{n} = f^{-}, \\\\
    f^{TB}(\eta, R)\equiv \frac{1}{\epsilon}\sum_{n,m,j,k=0}^{\infty}\zeta_{n,j}R^{j}\ \lambda_{m,k}\epsilon^{k},\\\\
    f^{BL}(\eta) \equiv \lim^{\mathcal{M}^{+}}_{\eta \to \epsilon(\Sigma)} \sum_{n=0}^{\infty}\left(-\frac{1}{2}\right)^{n}\frac{A_{n}\sigma^{n+1}}{\left(3n + 2\right)!}\eta^{3n+2} = f^{+},
\end{array}
\end{equation}
\\
where $\epsilon(\Sigma)$ denotes the value of $\epsilon$ at $\Sigma$, the boundary of the two set spaces and $f^{TB}$ stands for the transition boundary solution. Since the parameter $\epsilon$ is valid only in the leading edge subset and since it's a means of measure of the size of the set, we can conclude that it approaches its maxima when evaluated at the boundary. Continuity of the solution across the boundary dictates the following \textit{matching condition},
\\
\begin{equation}
    f^{+} = f^{-}.
\end{equation}

\begin{theorem}[Preliminary Matching Condition]
Let $\mathcal{M}^{+}$, $\mathcal{M}^{-}$ and $\Sigma$ denote the set of solutions in the trailing edge, the set of solutions in the leading edge and the boundary surface separating the two sets respectively. Let a parameter $\epsilon$ limit the size of $\mathcal{M}^{-}$ such that $max(\epsilon) \equiv \epsilon(\Sigma)$. For the real-valued functions $f^{-}: \mathcal{M}^{-}\rightarrow \mathbb{R}$ and $f^{+}: \mathcal{M}^{+}\rightarrow \mathbb{R}$ that have supports in $[0,\epsilon)$ and $(\epsilon, \infty)$ respectively, the preliminary matching condition that dictates continuity in the global solution across the boundary reads
\\
\begin{equation}
    [f]\equiv f^{+}-f^{-} = 0.
\end{equation}
\end{theorem}
\noindent Since the solution $f(\eta,R)$ at the leading obeys the governing PDE which in turn was obtained from the pressure-less differential equation in terms of the stream function which in turn was obtained from the continuity and momentum equations, the continuity of the solution across the junction implies that the conservation equations hold.

\subsection{Momentum ODE-Blasius Boundary Matching}
\noindent We consider the matching between the momentum ODE at the leading edge and the inner-transition boundary solutions. Since the Painlev\'e solution of the momentum ODE consisted of four terms as given in \ref{eq:finalsolPainleveODE}, matching these solutions give us
\\
\begin{equation}
    \lim_{\eta\to\epsilon(\Sigma)}\left(\frac{3\left(1+\eta_{0} \right)}{R\left(\eta - \eta_{0} \right)} + \frac{4\left(\eta_{0} - 2\eta_{0}^{3} - \eta_{0}^{5}\right)}{R\left(1+\eta^{2} \right)^{2}} + \frac{2\left(1 + 19\eta_{0}^{2} + 3\eta_{0}^{4} + \eta_{0}^{6} \right)\left(\eta - \eta_{0} \right)}{R\left(1+\eta^{2} \right)^{3}}\right) = \frac{1}{\epsilon}\sum_{k=0}^{3}\lambda^{-}_{k}\epsilon^{k},
\end{equation}
\\
where the transition boundary solution has been written with a simple pole-type singularity. Taking the limit and Taylor expanding, we obtain the following relations
\\
\begin{equation}\label{matchedLE-TS}
    \lambda^{-}_{0} =\frac{3\left(1+\eta_{0} \right)}{R},\ \lambda^{-}_{1} = \lambda^{-}_{2} = \lambda^{-}_{3} = 0.
\end{equation}
\\
Matching the Blasius solution to the outer-transition boundary solution, we list the non-zero constants post the matching 
\\
\begin{equation}\label{matchedBL-TS}
    \begin{array}{lr}
    \lim_{\eta\to\epsilon(\Sigma)}\sum_{n=0}^{\infty}\left(-\frac{1}{2}\right)^{n}\frac{A_{n}\sigma^{n+1}}{\left(3n + 2\right)!}\eta^{3n+2} = \frac{1}{\epsilon}\sum_{k=0}^{\infty}\lambda^{+}_{k}\epsilon^{k},\\\\
    
    \lambda^{+}_{3} = \frac{1}{2}\sigma,\ \lambda^{+}_{9} = \frac{1}{4}\frac{A_{2}\sigma^{3}}{8!},\ \lambda^{+}_{12} = -\frac{1}{8}\frac{A_{3}\sigma^{4}}{11!},\ldots
    \end{array}
\end{equation}
\\
From equations \ref{matchedLE-TS} and \ref{matchedBL-TS} we obtain the transition solution valid on $\Sigma$ such that
\\
\begin{equation}
    \left[\lambda_{i}\right] = \lambda^{+}_{i}-\lambda^{-}_{i},
\end{equation}
\\
and using this, we obtain,
\\
\large{
\begin{equation}
   \begin{array}{lr}
    f^{TB}(\eta)|_{\eta = \epsilon} = \frac{1}{\epsilon}\left[\lambda_{0}\right] + \left[\lambda_{3}\right]\epsilon^{2} + \left[\lambda_{9}\right]\epsilon^{8} + \left[\lambda_{12}\right]\epsilon^{11} + \ldots\\\\
    \ \ \ \ \ \ \ \ \ \ \ \ \ \ = -\frac{3\left(1+\eta_{0}\right)}{R\epsilon} + \frac{1}{2}\sigma\epsilon^{2} - \frac{1}{4}\frac{A_{2}\sigma^{3}}{8!}\epsilon^{8} -\frac{1}{8}\frac{A_{3}\sigma^{4}}{11!}\epsilon^{11} + \ldots\\\\
    \ \ \ \ \ \ \ \ \ \ \ \ \ \ = -\frac{3\left(1+\eta_{0}\right)}{R\epsilon} + \sum_{n=0}^{\infty}\left(-\frac{1}{2}\right)^{n}\frac{A_{n}\sigma^{n+1}}{\left(3n+2\right)!}\epsilon^{3n+2}.
   \end{array}
\end{equation}}
\\
It can be observed from this equation that far downstream, for larger Reynolds number, the first term vanishes and we are left with the power series solution of the Blasius equation.

\subsection{Momentum PDE-Blasius Boundary Matching}
\noindent The preliminary junction condition will fix the constraints on the coefficients, $\alpha_{n}(R)$ and $\beta_{m}(\eta)$, when they are matched with the leading edge and the trailing edge solution at $\Sigma$. We first consider the matching between the leading edge and the transition boundary solutions. Since the Painlev\'e solution of the momentum PDE consisted of four terms (since the highest positive Fuch's index was four as in \ref{eq:Polynomialequation}), we perform the summation up to only four indices and this gives
\\
\begin{equation}
   \lim_{\eta \to \epsilon(\Sigma)}\left( A_{0}\xi^{-1} + A_{1} + A_{2}\xi + A_{3}\xi^{2}\right) = \frac{1}{\epsilon}\sum_{m,n,j,k=0}^{3}\zeta_{n,j}R^{j}\ \lambda_{m,k}\eta^{k},
\end{equation}
\\
where $A_{3} = 0$ and is the coefficient of the Kovalevskaya exponent $\xi^{2} = \left(\eta - \eta_{0}(R)\right)^{2}$. Performing the calculations, similar to the one done in the previous section, we obtain the following transition solution
\\
\begin{equation}
    f^{TB}(\eta,R)|_{\eta=\epsilon} = \frac{3\left(1+\eta_{0}^{2} - 2R\eta_{0}\eta_{0}' + R^{2}\left(\eta_{0}'\right)^{2} \right)^{2}}{R\left(1+\eta_{0}^{2}\right)\epsilon}  + \sum_{n=0}^{\infty}\left(-\frac{1}{2}\right)^{n}\frac{A_{n}\sigma^{n+1}}{\left(3n+2\right)!}\epsilon^{3n+2}.
\end{equation}
\\
This transition solution provides us insights into the role played by the derivative of $\eta_{0}(R)$. Note that this solution will simplify to the Blasius solution in the large limit of Reynolds number if and only if the derivative $\eta_{0}'(R)$ vanishes, i.e., $\eta_{0}'(R) = 0$. Although we do not know the form of $\eta_{0}(R)$, we now know that the function is to obey the condition that its first derivative is to vanish. The implication of this is twofold- one, in order to prevent a blow up and two, in order for the trailing edge solution to hold. Thus, the transition solution consists of two parts, one which contains the $R$-dependence and another independent of the same which further downstream (when $\eta>\epsilon(\Sigma)$) in the large Reynolds number limit (as the former falls off as $1/R$) results in the Blasius solution, i.e,
\\
\begin{equation}
    \lim_{R \to \infty}f^{TB}(\eta,R) = f^{BL}(\eta),\ iff \ \eta_{0}'(R)=0.
\end{equation}

\section{The Homotopy Perturbation Method}
\noindent The Homotopy-perturbation method is a new type of perturbation method proposed by He \cite{He}. We also closely follow the methodology given in \cite{He2} and \cite{He3} to construct a solution for the fourth-order non-linear self-similar momentum equation. This method, in contrast to the other general perturbation ones, does not call for a dependence of a small parameter in the equation. 
\newline
A general non-linear PDE is considered of the following type
\\
\begin{equation}\label{eq:mainhomotopy}
    \mathcal{D}(x) - f(r) = 0,\ r\in \mathcal{M},
\end{equation}
\\
with boundary conditions 
\\
\begin{equation}
    \mathcal{B}\left(u, \frac{\partial u}{\partial n} \right) = 0,\ r\in \partial \mathcal{M},
\end{equation}
\\
where $\mathcal{D}$ is a differential operator, $\mathcal{B}$ is a boundary operator and $f(r)$ is an analytic function that is prescribed, and $\partial \mathcal{M}$ is the boundary of the domain $\mathcal{M}$. We now divide the differential operator $\mathcal{D}$ into linear ($\mathcal{L}$) and non-linear ($\mathcal{N}$) as follows
\\
\begin{equation}
    \mathcal{L}[x] - \mathcal{N}[x] - f(r) = 0.
\end{equation}
\\
Upon embedding an artificial parameter $p$ in the non-linear equation, the equation reads
\\
\begin{equation}
    \mathcal{L}[x] - p\mathcal{N}[x] - f(r) = 0,
\end{equation}
\\
such that the approximate solution to the differential equation can be expressed as the following power series expansion in terms of $p$
\\
\begin{equation}
    x = x_{0} + px_{1} + p^{2}x_{2} + ... = \sum_{i=0}^{n}p^{i}x_{i}.
\end{equation}
\\
Now, in the limit of $p\rightarrow 1$, the above series becomes the approximate solution to the original differential equation. Although this method yields good approximate solutions for certain classes of non-linear differential equations, the arbitrariness associated with the embedding parameter $p$ leads to non-uniform approximations. 
\newline
Homotopy, a key topic in differential topology, was made use by Liao [\cite{Liao1},\cite{Liao2}] to construct this perturbation method. The first step in this method is to construct a homotopy of the differential equation $y(r,p):\mathcal{M}\times [0,1]\rightarrow \mathbb{R}$ which satisfies
\\
\begin{equation}\label{eq:referencehomo}
    \mathcal{W}(y,p) \equiv (1-p)\left(\mathcal{L}[y] - \mathcal{L}[x_{0}]\right) + p\left(\mathcal{D}[y] - f(r) \right) = 0,\ p\in [0,1],\ r\in \mathcal{M},
\end{equation}
\\
which when simplified reads
\\
\begin{equation}\label{eq:homotopy}
     \mathcal{W}(y,p) \equiv \mathcal{L}[y] - \mathcal{L}[x_{0}] + p\mathcal{L}[x_{0}] + p\left(\mathcal{N}[y] - f(r) \right) = 0,
\end{equation}
\\
where $x_{0}$ in an initial approximation which satisfies the prescribed boundary conditions. The solution to the \ref{eq:homotopy} can be expressed as follows
\\
\begin{equation}\label{eq:HPMseries}
    y = \sum_{i=0}^{n}p^{i}y_{i},
\end{equation}
\\
and hence, the approximate solution to \ref{eq:mainhomotopy} can be obtain by limiting the value of $p$ to unity in the solution to $y$, i.e.,
\\
\begin{equation}
    x = \lim_{p\to 1}y = \sum_{i=0}^{n}y_{i}.
\end{equation}
\\
It is to be noted here that the above series solution does indeed converge as proven in \cite{convergence}.

\section{Application to the Blasius Equation}\label{sec:BlasiusHPM}
\noindent The linear and non-linear parts of the Blasius equation are identified to be
\\
\begin{equation}
    \mathcal{L}[\cdot] = \frac{d^{3}}{d\eta^{3}}[\cdot],\ \mathcal{N}[\cdot] = \frac{1}{2}[\cdot]\frac{d^{2}}{d\eta^{2}}[\cdot],
\end{equation}
\\
which when input into \ref{eq:referencehomo} reads
\\
\begin{equation}\label{Blasiushomo}
     \mathcal{W}(f,p) = (1-p)\left(\mathcal{L}[f] - \mathcal{L}[g]\right)  + p\left(\mathcal{N}[f] + \mathcal{L}[f] \right) = 0.
\end{equation}
\\
Now, we take initial approximation to be the following
\\
\begin{equation}
    g(\eta) = \frac{2}{\pi}tan^{-1}\left(\frac{\pi}{2}\eta \right),
\end{equation}
\\
and we also note that the above approximation obeys all the prescribed boundary conditions of the Blasius equation\footnote{i.e., $g(0)=g'(0)=0$ and $g'(\infty) = 1$}. We now express $f$ and its derivatives as a series of the powers of $p$ as follows
\\
\begin{equation}
\begin{array}{lr}

    f = f_{0} + pf_{1} + ... = \sum_{i=0}^{n}p^{i}f_{i},\\\\
    f'' = \sum_{i=0}^{n}p^{i}f''_{i},\\\\
    f''' = \sum_{i=0}^{n}p^{i}f'''_{i},\\\\
\end{array}
\end{equation}
\\
which when substituted into \ref{Blasiushomo} reads
\\
\begin{equation}
     \sum_{i=0}^{n}p^{i}f'''_{i} - (1+p)\frac{2}{\pi}tan^{-1}\left(\frac{\pi}{2}\eta \right) - p\sum_{i=0}^{n}p^{i}f'''_{i} + p\sum_{i=0}^{n}p^{i}f_{i}\sum_{i=0}^{n}p^{i}f''_{i} + p\sum_{i=0}^{n}p^{i}f'''_{i} = 0.
\end{equation}
\\
Here, we note that around the value of the self-similar variable when at boundary layer thickness ($\delta$), $\eta = \eta_{(\delta)}=\Delta$, we have by definition $u/U_{\infty} \approx 0.99$, where $U_{\infty}$ is the free-stream velocity. Thus, instead of setting $\eta=\infty$ at the far-field, we set $\eta = \Delta$ which replaces the boundary condition $f'(\infty) = 1$ with $f'(\Delta)= 1$. Now, we append the initial approximation as follows to suit the modified boundary condition and set it equal to the solution of the zeroth-order equation\footnote{The value of $\Delta$ is found to be $8.6989$ via numerical solving the ODE.}, i.e., 
\\
\begin{equation}
    g(\eta) = f_{0}(\eta) = \frac{2\eta^{2}}{\Delta\left(1+\frac{\eta^{2}}{\Delta^{2}}\right)}.
\end{equation}
\\
Limiting to $\mathcal{O}(p^{2})$ for the sake of simplicity and matching powers of $p$ yields
\\
\begin{equation}
    \begin{array}{lr}
    p^{0}:\ f'''_{0} - g''' = 0;\ f_{0}(0) = f_{0}'(0) = 0\   \&\  f'_{0}(\Delta)= 1\\\\
    p^{1}:\ f_{1}''' + g''' + \frac{1}{2}f_{0}f''_{0} = 0;\ f_{1}(0) = f'_{1}(0) = 0\  \&\ f'_{1}(\Delta)= 1 \\\\
    p^{2}:\ f'''_{2} + \frac{1}{2}f_{0}f''_{1} + \frac{1}{2}f_{1}f''_{0} = 0,\ f_{2}(0) = f'_{2}(0) = 0\ \&\ f'_{2}(\Delta)= 1.
    \end{array}
\end{equation}
\\
Thus, the solution to the ODE obtain by matching of coefficients of the powers of $p$ reads
\\
\begin{equation}
\begin{array}{lr}
    f_{0} = \frac{2x^{2}}{\Delta\left(1+\frac{x^{2}}{\Delta^{2}}\right)} \approx 0.229914\ \eta^{2} - 0.00303834\ \eta^{4} + 0.0000401521\ \eta^{6}\\\\
    \ \ \ \ \ \ \ \ \ \ \ \ \ \ \ \ \ \ \ \ - 5.30615\times 10^{-7}\ \eta^{8} + 7.01215\times 10^{-9}\ \eta^{10} + \mathcal{O}(\eta^{11}),\\\\
   
    f_{1} = -\frac{7 \Delta^4 \eta}{4 (\Delta^2 + \eta^2)} - \frac{3 \Delta \eta^2}{2 (\Delta^2 + \eta^2)} + \frac{5 \Delta^3 \eta^2}{24 (\Delta^2 + \eta^2)} - \frac{\pi \Delta^3 \eta^2}{16 (\Delta^2 + \eta^2)} - \frac{17 \Delta^2 \eta^3}{12 (\Delta^2 + \eta^2)}\\\\
 \ \ \ \ \ \ + \frac{\eta^4}{2 \Delta (\Delta^2 + \eta^2)} + \frac{5 \Delta \eta^4}{24 (\Delta^2 + \eta^2)} - \frac{\pi \Delta \eta^4}{16 (\Delta^2 + \eta^2)} + \frac{
 7 \Delta^5 tan^{-1}\left(\frac{\eta}{\Delta} \right)}{4 (\Delta^2 + \eta^2)} + \frac{2 \Delta^3 \eta^2  tan^{-1}\left(\frac{\eta}{\Delta} \right)}{(\Delta^2 + \
\eta^2)} + \frac{\Delta \eta^4  tan^{-1}\left(\frac{\eta}{\Delta} \right)}{4 (\Delta^2 + \eta^2)}\\\\
 \ \ \ \ \ \approx -0.0681898\ \eta^2 + 0.00303834\ \eta^4 - 0.000881008\ \eta^5 - 
 0.0000401521\ \eta^6\\\\
 \ \ \ \ \ \ +  0.0000232853\ \eta^7 + 5.30615\times10^{-7}\ \eta^8 - 4.02964\times10^{-7}\ \eta^9 -\\\\
 \ \ \ \ \ \ - 7.01215\times10^{-9}\ \eta^{10} + \mathcal{O}(\eta^{11}).
\end{array}
\end{equation}
\\
The solution to the second-order equation is highly complicated and this arises from our choice of the test function. The advantage of the modified boundary condition is that it allows for a wider class of test functions. The final solution can be constructed to be 
\\
\begin{equation}\label{approxBlasius}
    f(\eta) \approx 0.161724\  \eta^2 - 0.000881008\ \eta^5 + 0.0000232853\ \eta^7 - 4.02964\times 10^{-7}\ \eta^9 .
\end{equation}
\\
There is another way to work around this- by taking the initial approximation to be equal to the function itself. By doing this there is no need to compromise the last boundary condition we have modified above. Setting the initial approximation to be equal to $f_{0}(\eta)$ and expanding the later in a Taylor series centered around $\eta=0$, we have  
\\
\begin{equation}\label{initialapproximation}
\begin{array}{lr}
    g(\eta) = f_{0}(\eta) = f(0) + f'(0)\eta + \frac{1}{2}f''(0)\eta^{2} + ...\\\\
    \ \ \ \ \ \ = 0.1660285\ \eta^{2}.
\end{array}
\end{equation}
\\
It is to be noted here that since we are expanding the function $f_{0}(\eta)$ in a Taylor series, the initial approximation gets Taylor expanded. The last line was obtained from the fact that for the Blasius equation we have $f''(0) = 0.332057$ \cite{Blasius}. Limiting to $\mathcal{O}(p^{2})$ and matching powers of $p$, we have the following set of equations
\\
\begin{equation}
    \begin{array}{lr}
    f_{0}'''-g'''=0;\ f_{0}(0)=f'_{0}(0)=0,\ \&\ f_{0}'(\infty)=1\\\\
    f'''_{1}+g'''+\frac{1}{2}f_{0}f_{0}''=0;\ f_{1}(0)=f'_{1}(0)=0,\ \&\ f_{1}'(\infty)=0\\\\
    f_{2}'''+\frac{1}{2}f_{1}f_{0}''+\frac{1}{2}f_{0}f_{1}''=0;\ f_{2}(0)=f'_{2}(0)=0,\ \&\ f_{2}'(\infty)=0.\\\\
    \end{array}
\end{equation}
Now, the solutions to the zeroth-order, first-order and second-order equations are 
\\
\begin{equation}
    \begin{array}{lr}
    f_{0} = 0.1660285\ \eta^{2},\\\\
    f_{1} = -0.000459424 \eta^{5},\\\\
    f_{2} = 2.49718\times10^{-6}\ \eta^{8}.
    \end{array}
\end{equation}
\\
These yield the following solution to the order $\mathcal{O}(p^{2})$
\\
\begin{equation}\label{HPMBlasius}
  f(\eta) = 0.1660285\ \eta^{2} - 0.000459424\ \eta^{5} + 2.49718\times10^{-6}\ \eta^{8}.  
\end{equation}
\\
We can compare the solutions \ref{approxBlasius} and \ref{HPMBlasius} to observe the difference in the two approaches- the former with a constructed initial approximation that obeys the boundary conditions and the latter with a Taylor expanded initial approximation whose coefficients are automatically fixed by the available boundary condition data. Table \ref{tab:BLHPMvNUM} compares the data obtained via HPM (solution \ref{HPMBlasius}) to numerical data.
\\
\begin{table}
    \centering
    \begin{tabular}{c c c c c}
        Sl. & $\eta$ & HPM & Numerical & Error (\%) \\ [0.5ex]
        \hline\hline
        $1.$  & $0.4$ & $0.026559$ & $0.02656$ & $0.00054$ \\\\
        $2.$  & $0.8$ & $0.106108$ & $0.10611$ & $0.00179$ \\\\
        $3.$  & $1.2$ & $0.237948$ & $0.23795$ & $0.00060$ \\\\
        $4.$  & $2.4$ & $0.92249$ & $0.92230$ & $-0.02068$ \\\\
        $5.$  & $3.6$ & $1.94438$ & $1.92954$ & $-0.76917$ \\[1ex] 
        \hline\hline
    \end{tabular}
    \caption{Comparison of the data obtained from the HPM solution to the numerical data from literature.}
    \label{tab:BLHPMvNUM}
\end{table}

\section{Application to the Falkner-Skan Equation}
\noindent The linear and non-linear parts of the Falkner-Sknan equation are identified to be 
\\
\begin{equation}\label{eq:FSLNLoperator}
    \mathcal{L}[\cdot] = \frac{d^{3}}{d\eta^{3}}[\cdot],\ \mathcal{N}[\cdot] = [\cdot]\frac{d^{2}}{d\eta^{2}}[\cdot] - \beta\left(\frac{d}{d\eta}[\cdot] \right)^{2},
\end{equation}
\\
and we consider the same initial approximation that was consider for the Blasius case (\ref{initialapproximation}). Limiting to $\mathcal{O}(p^{2})$ and matching powers of $p$, we have the following set of equations
\\
\begin{equation}
    \begin{array}{lr}
    p^{0}:\ f_{0}'''-g'''=0;\ f_{0}(0)=f'_{0}(0)=0,\ \&\ f_{0}'(\Delta)=1\\\\
    p^{1}:\ f'''_{1} + f_{0}f_{0}'' + \beta\left(1 - \left( f_{0}'\right)^{2} \right)=0;\ f_{1}(0)=f'_{1}(0)=0,\ \&\ f_{1}'(\Delta)=0\\\\
    p^{2}:\ f_{2}''' - f_{1}''' + f_{0}f_{1}'' + f_{1}f_{0}'' - \beta\left(f_{1}' \right)^{2} = 0;\ f_{2}(0)=f'_{2}(0)=0,\ \&\ f_{2}'(\Delta)=0.\\\\
    \end{array}
\end{equation}
\\
Now, calculations of $f_{i}(\eta)$ will depend upon the value of $\beta$ and for each value of the same there would be a different lower limit for $\Delta$ above which the value of $f''(0)$ remains unchanged.
\newline
\newline
a. Case I: $\beta = 0$
\newline
We have here $f''(0) = 0.4696$, which  was unchanged for values of $\Delta\geq 8.6989$. Thus, the solutions read
\\
\begin{equation}
    \begin{array}{lr}
    f_{0} = 0.2348\eta^{2},\\\\
    f_{1} = - 0.0018377 \eta^5,\\\\
    f_{2} = 0.0000282525 \eta^8.
    \end{array}
\end{equation}
\\
Thus, the solution in this case is
\\
\begin{equation}\label{beta=0}
  f(\eta) =    0.2348\eta^{2} - 0.0018377 \eta^5 + 0.0000282525 \eta^8.
\end{equation}
\\
b. Case II: $\beta = -0.1988$
\newline
We have here $f''(0) = 0.00521828$, which  was unchanged for values of $\Delta\geq 8.6989$. Thus, the solutions read
\\
\begin{equation}
    \begin{array}{lr}
    f_{0} = 0.00260914\ \eta^{2},\\\\
    f_{1} = -2.2692\times 10^{-7}\ \eta^5,\\\\
    f_{2} = 3.87663\times 10^{-11}\ \eta^8.
    \end{array}
\end{equation}
\\
Thus, the solution in this case is
\\
\begin{equation}\label{beta=-0.1988}
  f(\eta) =   0.00260914\ \eta^{2} - 2.2692\times 10^{-7}\ \eta^5 +  3.87663\times 10^{-11}\ \eta^8.
\end{equation}
\\
c. Case III: $\beta = -0.1$
\newline
We have here $f''(0) = 0.31927$, which  was unchanged for values of $\Delta\geq 6.4$. Thus, the solutions read
\\
\begin{equation}\label{beta=-0.1}
    \begin{array}{lr}
    f_{0} = 0.159635\ \eta^{2},\\\\
    f_{1} = -0.000849446\ \eta^5,\\\\
    f_{2} = 8.87866\times 10^{-6}\ \eta^8.
    \end{array}
\end{equation}
\\
Thus, the solution in this case is
\\
\begin{equation}
  f(\eta) =   0.159635\ \eta^{2} - 0.000849446\ \eta^5 + 8.87866*10^-6\times 10^{-6}\ \eta^8.
\end{equation}
\\

\section{Application to the Self-Similar Momentum ODE}
\noindent The self-similar momentum ODE ($f\rightarrow f(\eta)$) defined at the leading edge reads
\\
\begin{equation}\label{HPMmomODE}
\begin{array}{lr}
    (1+\eta^{2})^{2}f'''' + 8\eta (1+\eta^{2})f''' + 4(1+3\eta^{2})f'' + R\left(2\eta ff'' + (1+\eta^{2})\left(ff'' \right)^{'} \right) = 0,
\end{array}
\end{equation}
\\
with boundary conditions $f(0)=f'(0) = 0$ at the plate (i.e., at $\eta=0$) and $f'(\infty)\rightarrow 1,\ f(\infty)-\eta f'(\infty)\rightarrow 0$ at far-field (i.e., at $\eta\to \infty$). The linear and non-linear parts of the ODE are identified to be
\\
\begin{equation}\label{HPMmomoperator}
\begin{array}{lr}
    \mathcal{L}[f] = f'''' + 4f''',\\\\
    \mathcal{N}[f] = \eta\left(8f''' + 2Rff'' \right) + \eta^{2}\left(2f'''' + 12f'' + R(ff'')^{'} \right) + 8\eta^{3}f''' + \eta^{4}f''''.
\end{array}    
\end{equation}
\\
We take the initial approximation to be 
\\
\begin{equation}\label{HPMinitialmomODE}
\begin{array}{lr}
    g(\eta) = f_{0}(\eta) = f(0) + f'(0)\eta + \frac{1}{2}f''(0)\eta^{2}\\\\
    \ \ \ \ \ \ = \frac{1}{2}f''(0)\eta^{2},
\end{array}
\end{equation}
\\
Now, to fix the value of $f''(0)$, we are to solve the ODE numerically for specific values of Reynolds number. Matching powers of $p$ yields
\\
\begin{equation}
    \begin{array}{lr}
    p^{0}:\ f_{0}'''' - g'''' +4\left(f_{0}'' - g'' \right) = 0;\ f_{0}(0)=f_{0}'(0)=0,\ f_{0}(\Delta) = f_{0}(\Delta)-\Delta f_{0}'(\Delta)=1\\\\
    p^{1}:\ f_{1}'''' + 4f_{1}'' - f_{0}'''' - 4f_{0}'' + g'''' + 4g'' + \eta\left(8f_{0}''' + 2Rf_{0}f_{0}'' \right) \\\\
    \ \ \ \ \ \ + \eta^{2}\left(2f_{0}'''' + 12f_{0}'' + R f_{0}'f_{0}'' + R f_{0}f_{0}'''' \right) + 8\eta^{3}f_{0}''' + \eta^{4}f_{0}'''' = 0;\ f_{1}(0)=f_{1}'(0)=0,\\\\
    \ \ \ \ \ \ f_{1}(\Delta) = f_{1}(\Delta)-\Delta f_{1}'(\Delta)=1\\\\
    p^{2}:\ f_{2}'''' + 4f_{2}'' - f_{1}'''' - 4f_{1}'' + \eta\left(8f_{1}''' + 2Rf_{0}f_{1}'' + 2Rf_{1}f_{0}'' \right)\\\\
    \ \ \ \ \ \ + \eta^{2}\left(2f_{1}'''' +  12f_{1}'' + R \left(f_{0}'f_{1}'' +f_{1}'f_{0}'' + f_{0}f_{1}''' + f_{1}f_{0}'''\right) \right) + 8\eta^{3}f_{1}''' + \eta^{4}f_{1}'''' = 0;\\\\
    \ \ \ \ \ \ f_{2}(0)=f_{2}'(0)=0,\ f_{2}(\Delta) = f_{2}(\Delta)-\Delta f_{2}'(\Delta)=1
    \end{array}
\end{equation}
\\
When the Reynolds number is set to unity, we have $f''(0)\approx 1.44657$. But since the values of $R$ are very small near the leading edge, we consider cases when $R=10^{-2}$, and $R=10^{-3}$ separately. We propose an algorithm in which we first need to fix the value of $\Delta$ by numerically solving the ODE such that the value of $f''(0)$ shows very small fluctuations for a value $\Delta \geq \kappa$. At the leading edge, to fix a domain, we set an upper limit on the value of $y$ and proceed to numerical trials for smaller and smaller values of $x$. Table \ref{tab:AllRf''} shows the values of $\Delta$ at which the corresponding $f''(0)$ shows least fluctuations for different cases of $R$. The numerical runs have been tabulated in appendix \ref{appendix1}. 
\\
\begin{table}
    \centering
    \begin{tabular}{c c c c}
        Sl. & $\Delta$ & $f''(0)$ & $R$ \\ [0.5ex]
        \hline\hline
        $1.$  & $999$ & $1.44657$ & $1$\\\\
        $2.$ & $886$ & $1.37435$ & $0.1$\\\\
        $3.$ & $\approx10^{5}$ & $1.36378$ & $0.01$\\\\
        $4.$ & $\approx10^{5}$ & $1.36306$ & $0.001$\\ [1ex] 
        \hline\hline
    \end{tabular}
    \caption{Value of $\Delta$ for which the corresponding $f''(0)$ shows minimal fluctuations for different cases of Reynolds number.}
    \label{tab:AllRf''}
\end{table}

\section{Application to the Self-Similar Momentum PDE}
\noindent Here we provide an overview of the steps to be followed to obtain the solution to the fourth-order non-linear self-similar momentum PDE up to terms of  $\mathcal{O}(R)$ (as was shown in \ref{momeqO1}) via HPM. Firstly, we identify the linear and non-linear operators,
\\
\begin{equation}\label{HPMmomPDEoperators}
    \begin{array}{lr}
    \mathcal{L}[\cdot] = \frac{d^{4}}{d\eta^{4}}[\cdot] + 4\left(\frac{d^{2}}{d\eta^{2}}[\cdot] - R\frac{d^{3}}{d\eta^{3}}[\cdot] \right),\\\\
    
    \mathcal{N}[\cdot] = \eta^{4}\frac{d^{4}}{d\eta^{4}}[\cdot] +  8\eta(1+\eta^{2})\frac{d^{3}}{d\eta^{3}}[\cdot] + \eta^{2}\left(12\frac{d^{2}}{d\eta^{2}}[\cdot] + 2\frac{d^{4}}{d\eta^{4}}[\cdot]\right)\\\\
    +R \left(2\eta[\cdot]\frac{d^{2}}{d\eta^{2}}[\cdot] +(1+\eta^{2})\left([\cdot]\frac{d^{3}}{d\eta^{3}}[\cdot] + \frac{d}{d\eta}[\cdot]\frac{d^{2}}{d\eta^{2}}[\cdot]\right) -12\eta^{2}\frac{d^{3}}{dRd\eta^{2}}[\cdot] - 4\eta(1+\eta^{2})\frac{d^{4}}{dRd\eta^{3}}[\cdot] \right).
    \end{array}
\end{equation}
\\
Similar, to what was done in the previous section, we consider the initial approximation $h(\eta,R)$ to be equal to $f_{0}(\eta,R)$ with the latter expressed as a two-dimensional Taylor expansion (to second-order) of $f(\eta,R)$ around the leading edge with $\eta = 0$ and $R=\alpha$ (where $\alpha$ is a small value since Reynolds numbers at the leading edge are usually of the order of $10^{-3}$), i.e.,
\\
\begin{equation}\label{HPMmomPDEinitial}
\begin{array}{lr}
    h(\eta,R) = f_{0}(\eta,R) = f(0,\alpha) + f_{\eta}(0,\alpha)\eta + f_{R}(0,\alpha)(R-\alpha)\\\\ 
    \ \ \ \ \ \ \ \ \ \ \ \ +\frac{1}{2}\left(f_{\eta\eta}(0,\alpha)\eta^{2} + 2f_{\eta R}(0,\alpha)\eta (R-\alpha) + f_{RR}(0,\alpha)(R-\alpha)^{2} \right)\\\\
    \ \ \ \ \ \ \ \ \ \  = \frac{1}{2}\left(f_{\eta\eta}(0,\alpha)\eta^{2} + 2f_{\eta R}(0,\alpha)\eta (R-\alpha) + f_{RR}(0,\alpha)(R-\alpha)^{2} \right) - \alpha f_{R}(0,\alpha).
\end{array}
\end{equation}
\\
The values of $f_{\eta\eta}(0,\alpha)$, $f_{\eta R}(0,\alpha)$, and $f_{RR}(0,\alpha)$ for a fixed value of $\alpha\  (>R)$ can be found via numerically solving the momentum PDE to  $\mathcal{O}(R)$.

\section{Conclusion}
\noindent In this paper, we have explored various topics in the leading edge problem with an emphasis on the self-similar solutions to the momentum and energy equations. The common theme that connects most of our work is the analysis of partial differential equations.  We have used a self-similar function, which has dependence on both the self-similar variable and the Reynolds number, to transform the energy and the momentum equations showing them to be second-order non-linear and fourth-order non-linear PDE respectively. Post the derivation of the energy PDE, the difficulty associated with solving the same has been discussed. We have proposed a boundary-matching technique and these solutions show how the dependence on Reynolds number falls off far downstream and yields the Blasius solution. We have also included a detailed discussion of semi-analytical solutions via the homotopy perturbation method. An algorithmic scheme which involves consideration of a multi-dimensional Taylor expansion as the initial approximation to the MPDE has been presented and the far-field value $\Delta$ has been found numerically for different cases of Reynolds number. 
\newline
\noindent There is a considerable amount of scope for future work to be done in the leading edge problem. A furthermore detailed investigation of the integrability of the MODE, MPDE, and the EPDE is required. Possible approaches and (quasi) algorithmic methods, apart from the Painlev\'e test, that can reveal interesting prospects could be to probe for the existence of Lax pairs, the inverse scattering transform, and the Hirota bilinear method. It is to be noted that although the MPDE passed the Painlev\'e test, we were unable to find a Lax pair. The implication of this could be that either the results are a bit deceiving since not all the equations that pass the Painlev\'e test are necessarily integrable-there is a very small subclass of counter examples-or that presently, we lack the mathematical competence a task such as this demands. To elaborate more on the former point, does the MPDE fall into that subclass? If so, how could we check for it? The prime question here would then be: What are some of the common features of the counterexamples and how do their solutions look like? Another key point where further thinking is required is the reason for the EPDE's non-integrability. Why is the MPDE integrable while the EPDE, whose solution depends upon the MPDE, fails the Painlev\'e test? These are some questions to ponder over and (hopefully) be resolved in future work(s). 

\section*{Acknowledgements}
\noindent The authors (Naveen Balaji, Sujan Kumar, and Vignesh) are greatly indebted to Dr. K N Seetharamu for his time, invaluable advice, for introducing them to the leading edge problem, and most importantly for tolerating their whims for close to a year. Naveen and Sujan would like to thank early collaborators which include Dr. Rammohan B for his deep insights and suggestions, Dr. T R Seetharamu for his deep insights and discussions, Mr. Babu Rao for his support, and Dr. K S Ravichandran for the long discussions and helpful suggestions. Without these generous men, our early foundational work would not have been possible. We would like to thank Dr. Vittal Rao of IISc., Bangalore for the multiple discussion sessions and Dr. Rodolfo Rosales of MIT, Boston for his deep insights and suggestions of literature.

\newpage
\section*{Appendix A}\label{appendix1}
\noindent Table \ref{tab:R=1} shows the results of the numerical runs for the case of $R=1$, table \ref{tab:R=0.1} shows the results of the numerical runs for the case of $R=0.1$, table \ref{tab:R=0.01} shows the results of the numerical runs for the case of $R=0.01$, table \ref{tab:R=0.001} shows the results of the numerical runs for the case of $R=0.001$. 

\begin{table}
    \centering
    \begin{tabular}{c c c c c}
        Sl. & $x$ & $y$ & $\eta=\Delta$ & $f''(0)$ \\ [0.5ex]
        \hline\hline
        $1.$ & $10^{-1}$ & $1$ & $10$ & $1.81176$\\\\
        $2.$ & $10^{-2}$ & $1$ & $100$ & $1.47821$\\\\
        $3.$ & $0.5\times 10^{-2}$ & $1$ & $200$ & $1.46063$\\\\
        $4.$ & $0.25\times 10^{-2}$ & $1$ & $400$ & $1.45185$\\\\
        $5.$ & $0.125\times 10^{-2}$ & $1$ & $800$ & $1.44745$\\\\
        $6.$ & $\frac{1}{9}\times 10^{-2}$ & $1$ & $900$ & $1.44696$\\\\
        $7.$ & $\frac{10}{99}\times 10^{-2}$ & $1$ & $990$ & $1.4466$\\\\
        $8.$ & $\frac{100}{999}\times 10^{-2}$ & $1$ & $999$ & $1.44657$\\ [1ex] 
        \hline\hline
    \end{tabular}
    \caption{Value of $f''(0)$ for $R=1$ and varying $\Delta$.}
    \label{tab:R=1}
\end{table}

\begin{table}
    \centering
    \begin{tabular}{c c c c c}
        Sl. & $x$ & $y$ & $\eta=\Delta$ & $f''(0)$ \\ [0.5ex]
        \hline\hline
        $1.$ & $10^{-1}$ & $1$ & $10$ & $1.69891$\\\\
        $2.$ & $10^{-2}$ & $1$ & $100$  & $1.40102$\\\\
        $3.$ & $0.5\times 10^{-2}$ & $1$ & $200$ & $1.38594$\\\\
        $4.$ & $0.25\times 10^{-2}$ & $1$ & $400$ & $1.37845$\\\\
        $5.$ & $0.125\times 10^{-2}$ & $1$  & $800$ & $1.37471$\\\\
        $6.$ & $\frac{5}{44}\times 10^{-2}$ & $1$ & $880$ & $1.37437$\\\\
        $7.$ & $\frac{50}{443}\times 10^{-2}$ & $1$ & $886$ & $1.37435$\\[1ex] 
        \hline\hline
    \end{tabular}
    \caption{Value of $f''(0)$ for $R=0.1$ and varying $\Delta$.}
    \label{tab:R=0.1}
\end{table}

\begin{table}
    \centering
    \begin{tabular}{c c c c c}
        Sl. & $x$ & $y$ & $\eta=\Delta$ & $f''(0)$ \\ [0.5ex]
        \hline\hline
        $1.$ & $10^{-1}$ & $1$ & $10$ & $1.68778$\\\\
        $2.$ & $10^{-2}$ & $1$ & $100$  & $1.3933$\\\\
        $3.$ & $0.5\times 10^{-2}$ & $1$ & $200$ & $1.37846$\\\\
        $4.$ & $0.25\times 10^{-2}$ & $1$ & $400$ & $1.37109$\\\\
        $5.$ & $0.125\times 10^{-2}$ & $1$  & $800$ & $1.36742$\\\\
        $6.$ & $\frac{1}{9}\times 10^{-2}$ & $1$ & $900$ & $1.36701$\\\\
        $7.$ & $\frac{10}{99}\times 10^{-2}$ & $1$ & $990$ & $1.36672$\\\\
        $8.$ & $10^{-3}$ & $1$ & $10^{3}$ & $1.36669$\\\\
        $9.$ & $10^{-4}$ & $1$ & $10^{4}$ & $1.36405$\\\\
        $10.$ & $10^{-5}$ & $1$ & $10^{5}$ & $1.36378$\\[1ex] 
        \hline\hline
    \end{tabular}
    \caption{Value of $f''(0)$ for $R=0.01$ and varying $\Delta$.}
    \label{tab:R=0.01}
\end{table}

\begin{table}
    \centering
    \begin{tabular}{c c c c c}
        Sl. & $x$ & $y$ & $\eta=\Delta$ & $f''(0)$ \\ [0.5ex]
        \hline\hline
        $1.$ & $10^{-1}$ & $1$ & $10$ & $1.68666$\\\\
        $2.$ & $10^{-2}$ & $1$ & $100$  & $1.39253$\\\\
        $3.$ & $0.5\times 10^{-2}$ & $1$ & $200$ & $1.37771$\\\\
        $4.$ & $0.25\times 10^{-2}$ & $1$ & $400$ & $1.37035$\\\\
        $5.$ & $0.125\times 10^{-2}$ & $1$  & $800$ & $1.36669$\\\\
        $6.$ & $\frac{1}{9}\times 10^{-2}$ & $1$ & $900$ & $1.36628$\\\\
        $7.$ & $\frac{10}{99}\times 10^{-2}$ & $1$ & $990$ & $1.36599$\\\\
        $8.$ & $10^{-3}$ & $1$ & $10^{3}$ & $1.36596$\\\\
        $9.$ & $10^{-4}$ & $1$ & $10^{4}$ & $1.36332$\\\\
        $10.$ & $10^{-5}$ & $1$ & $10^{5}$ & $1.36306$\\[1ex] 
        \hline\hline
    \end{tabular}
    \caption{Value of $f''(0)$ for $R=0.001$ and varying $\Delta$.}
    \label{tab:R=0.001}
\end{table}

\newpage
\newpage

\section*{Appendix B}\label{appendix2}
Below are the Lie prolongation terms. 
\begin{equation*}
\begin{array}{lr}
{\kappa }^{\mathrm{\etaup }\mathrm{\etaup }\mathrm{RR}}=\zeta f_{\mathrm{\etaup }\mathrm{\etaup }\mathrm{RR}}-\zeta f_{\mathrm{\etaup }\mathrm{\etaup }\mathrm{RRR}}+\xi f_{\mathrm{\etaup }\mathrm{\etaup }\mathrm{\etaup }\mathrm{R}}-\xi f_{\mathrm{\etaup }\mathrm{\etaup }\mathrm{\etaup }\mathrm{RR}}-6f_{\mathrm{\etaup }\mathrm{R}}f_{\mathrm{\etaup }\mathrm{RR}}{\zeta }_f-2f_{\eta }f_{\mathrm{\etaup }\mathrm{RRR}}{\zeta }_f-f_{\mathrm{RRR}}f_{\mathrm{\etaup }\mathrm{\etaup }}{\zeta }_f\\\\
\ \ \ \ \ \ \ \ \ \ \ -3f_{\mathrm{RR}}f_{\mathrm{\etaup }\mathrm{\etaup }\mathrm{R}}{\zeta }_f -3f_Rf_{\mathrm{\etaup }\mathrm{\etaup }\mathrm{RR}}{\zeta}_f-f_{\mathrm{RRR}}f^2_{\eta }{\zeta }_{\mathrm{ff}}-6f_{\mathrm{RR}}f_{\eta}f_{\mathrm{\etaup}\mathrm{R}}{\zeta}_{\mathrm{ff}}-6f_Rf^2_{\mathrm{\etaup }\mathrm{R}}{\zeta}_{\mathrm{ff}}-6f_Rf_{\eta}f_{\mathrm{\etaup }\mathrm{RR}}{\zeta}_{\mathrm{ff}}\\\\
\ \ \ \ \ \ \ \ \ \ \ -3f_Rf_{\mathrm{RR}}f_{\mathrm{\etaup}\mathrm{\etaup}}{\zeta}_{\mathrm{ff}}-3f^2_Rf_{\mathrm{\etaup}\mathrm{\etaup}\mathrm{R}}{\zeta }_{\mathrm{ff}}-3f_Rf_{\mathrm{RR}}f^2_{\eta}{\zeta}_{\mathrm{fff}}-6f^2_Rf_{\eta }f_{\mathrm{\etaup}\mathrm{R}}{\zeta }_{\mathrm{fff}}-f^3_Rf_{\mathrm{\etaup }\mathrm{\etaup}}{\zeta }_{\mathrm{fff}}-f^3_Rf^2_{\eta}{\zeta}_{\mathrm{ffff}}\\\\
\ \ \ \ \ \ \ \ \ \ \ -2f_{\mathrm{\etaup }\mathrm{\etaup }\mathrm{RR}}{\zeta }_R-4f^2_{\mathrm{\etaup }\mathrm{R}}{\zeta }_{\mathrm{Rf}}-4f_{\eta }f_{\mathrm{\etaup }\mathrm{RR}}{\zeta }_{\mathrm{Rf}}-2f_{\mathrm{RR}}f_{\mathrm{\etaup }\mathrm{\etaup }}{\zeta }_{\mathrm{Rf}}-4f_Rf_{\mathrm{\etaup }\mathrm{\etaup }\mathrm{R}}{\zeta}_{\mathrm{Rf}}-2f_{\mathrm{RR}}f^2_{\eta }{\zeta }_{\mathrm{Rff}}\\\\
\ \ \ \ \ \ \ \ \ \ \ -8f_Rf_{\eta }f_{\mathrm{\etaup }\mathrm{R}}{\zeta }_{\mathrm{Rff}}-2f^2_Rf_{\mathrm{\etaup }\mathrm{\etaup }}{\zeta }_{\mathrm{Rff}}-2f^2_Rf^2_{\eta }{\zeta }_{\mathrm{Rfff}}-f_{\mathrm{\etaup }\mathrm{\etaup }\mathrm{R}}{\zeta }_{\mathrm{RR}}-2f_{\eta }f_{\mathrm{\etaup }\mathrm{R}}{\zeta }_{\mathrm{RRf}}-f_Rf_{\mathrm{\etaup }\mathrm{\etaup }}{\zeta }_{\mathrm{RRf}}\\\\
\ \ \ \ \ \ \ \ \ \ \ -f_Rf^2_{\eta }{\zeta }_{\mathrm{RRff}}-2f_{\mathrm{\etaup}\mathrm{RRR}}{\zeta }_{\eta }-2f_{\mathrm{RRR}}f_{\eta }{\zeta }_{\mathrm{\etaup }\mathrm{f}}-6f_{\mathrm{RR}}f_{\mathrm{\etaup }\mathrm{R}}{\zeta }_{\mathrm{\etaup }\mathrm{f}}-6f_Rf_{\mathrm{\etaup }\mathrm{RR}}{\zeta }_{\mathrm{\etaup }\mathrm{f}}-6f_Rf_{\mathrm{RR}}f_{\eta }{\zeta }_{\mathrm{\etaup }\mathrm{ff}}\\\\
\ \ \ \ \ \ \ \ \ \ \ -6f^2_Rf_{\mathrm{\etaup }\mathrm{R}}{\zeta }_{\mathrm{\etaup}\mathrm{ff}}-2f^3_Rf_{\eta }{\zeta }_{\mathrm{\etaup }\mathrm{fff}}-4f_{\mathrm{\etaup }\mathrm{RR}}{\zeta}_{\mathrm{\etaup}\mathrm{R}}-4f_{\mathrm{RR}}f_{\eta }{\zeta}_{\mathrm{\etaup}\mathrm{Rf}}-8f_Rf_{\mathrm{\etaup }\mathrm{R}}{\zeta}_{\mathrm{\etaup}\mathrm{Rf}}-4f^2_Rf_{\eta }{\zeta }_{\mathrm{\etaup}\mathrm{Rff}}\\\\
\ \ \ \ \ \ \ \ \ \ \ -2f_{\mathrm{\etaup }\mathrm{R}}{\zeta }_{\mathrm{\etaup }\mathrm{RR}}-2f_Rf_{\eta }{\zeta }_{\mathrm{\etaup }\mathrm{RRf}}-f_{\mathrm{RRR}}{\zeta }_{\mathrm{\etaup }\mathrm{\etaup }}-3f_Rf_{\mathrm{RR}}{\zeta }_{\mathrm{\etaup }\mathrm{\etaup }\mathrm{f}}-f^3_R{\zeta }_{\mathrm{\etaup }\mathrm{\etaup }\mathrm{ff}}-2f_{\mathrm{RR}}{\zeta }_{\mathrm{\etaup }\mathrm{\etaup }\mathrm{R}}-2f^2_R{\zeta }_{\mathrm{\etaup }\mathrm{\etaup }\mathrm{Rf}}\\\\
\ \ \ \ \ \ \ \ \ \ \ -f_R{\zeta }_{\mathrm{\etaup }\mathrm{\etaup }\mathrm{RR}}+f_{\mathrm{\etaup }\mathrm{\etaup }\mathrm{RR}}{\kappa }_f+2f^2_{\mathrm{\etaup }\mathrm{R}}{\kappa }_{\mathrm{ff}}+2f_{\eta }f_{\mathrm{\etaup }\mathrm{RR}}{\kappa}_{\mathrm{ff}}+f_{\mathrm{RR}}f_{\mathrm{\etaup }\mathrm{\etaup}}{\kappa}_{\mathrm{ff}}+2f_Rf_{\mathrm{\etaup }\mathrm{\etaup}\mathrm{R}}{\kappa}_{\mathrm{ff}}+f_{\mathrm{RR}}f^2_{\eta }{\kappa }_{\mathrm{fff}}\\\\
\ \ \ \ \ \ \ \ \ \ \ +4f_Rf_{\eta }f_{\mathrm{\etaup }\mathrm{R}}{\kappa }_{\mathrm{fff}}+f^2_Rf_{\mathrm{\etaup }\mathrm{\etaup }}{\kappa }_{\mathrm{fff}}+f^2_Rf^2_{\eta }{\kappa }_{\mathrm{ffff}}+2f_{\mathrm{\etaup }\mathrm{\etaup }\mathrm{R}}{\kappa }_{\mathrm{Rf}} +4f_{\eta }f_{\mathrm{\etaup }\mathrm{R}}{\kappa }_{\mathrm{Rff}}+2f_Rf_{\mathrm{\etaup }\mathrm{\etaup }}{\kappa }_{\mathrm{Rff}}\\\\
\ \ \ \ \ \ \ \ \ \ \ +2f_Rf^2_{\eta }{\kappa }_{\mathrm{Rfff}}+f_{\mathrm{\etaup }\mathrm{\etaup }}{\kappa }_{\mathrm{RRf}}+f^2_{\eta }{\kappa }_{\mathrm{RRff}}+2f_{\mathrm{\etaup }\mathrm{RR}}{\kappa }_{\mathrm{\etaup }\mathrm{f}}+2f_{\mathrm{RR}}f_{\eta }{\kappa }_{\mathrm{\etaup }\mathrm{ff}}+4f_Rf_{\mathrm{\etaup }\mathrm{R}}{\kappa }_{\mathrm{\etaup }\mathrm{ff}} +2f^2_Rf_{\eta }{\kappa }_{\mathrm{\etaup }\mathrm{fff}}\\\\
\ \ \ \ \ \ \ \ \ \ \ +4f_{\mathrm{\etaup }\mathrm{R}}{\kappa }_{\mathrm{\etaup }\mathrm{Rf}} +4f_Rf_{\eta }{\kappa }_{\mathrm{\etaup }\mathrm{Rff}}+2f_{\eta }{\kappa }_{\mathrm{\etaup }\mathrm{RRf}}+f_{\mathrm{RR}}{\kappa }_{\mathrm{\etaup }\mathrm{\etaup }\mathrm{f}}+f^2_R{\kappa }_{\mathrm{\etaup }\mathrm{\etaup }\mathrm{ff}}+2f_R{\kappa }_{\mathrm{\etaup }\mathrm{\etaup }\mathrm{Rf}}+{\kappa }_{\mathrm{\etaup }\mathrm{\etaup }\mathrm{RR}}\\\\
\ \ \ \ \ \ \ \ \ \ \ -3f_{\mathrm{\etaup }\mathrm{RR}}f_{\mathrm{\etaup }\mathrm{\etaup }}{\xi}_f-6f_{\mathrm{\etaup }\mathrm{R}}f_{\mathrm{\etaup }\mathrm{\etaup }\mathrm{R}}{\xi }_f-3f_{\eta }f_{\mathrm{\etaup }\mathrm{\etaup }\mathrm{RR}}{\xi }_f-f_{\mathrm{RR}}f_{\mathrm{\etaup }\mathrm{\etaup }\mathrm{\etaup }}{\xi }_f-2f_Rf_{\mathrm{\etaup }\mathrm{\etaup }\mathrm{\etaup }\mathrm{R}}{\xi }_f-6f_{\eta }f^2_{\mathrm{\etaup }\mathrm{R}}{\xi }_{\mathrm{ff}}\\\\
\ \ \ \ \ \ \ \ \ \ \ -3f^2_{\eta }f_{\mathrm{\etaup }\mathrm{RR}}{\xi }_{\mathrm{ff}}-3f_{\mathrm{RR}}f_{\eta }f_{\mathrm{\etaup }\mathrm{\etaup }}{\xi }_{\mathrm{ff}} -6f_Rf_{\mathrm{\etaup }\mathrm{R}}f_{\mathrm{\etaup }\mathrm{\etaup }}{\xi }_{\mathrm{ff}}-6f_Rf_{\eta }f_{\mathrm{\etaup }\mathrm{\etaup }\mathrm{R}}{\xi }_{\mathrm{ff}}-f^2_Rf_{\mathrm{\etaup }\mathrm{\etaup }\mathrm{\etaup }}{\xi }_{\mathrm{ff}}-f_{\mathrm{RR}}f^3_{\eta }{\xi }_{\mathrm{fff}}\\\\
\ \ \ \ \ \ \ \ \ \ \ -6f_Rf^2_{\eta }f_{\mathrm{\etaup }\mathrm{R}}{\xi }_{\mathrm{fff}}-3f^2_Rf_{\eta }f_{\mathrm{\etaup }\mathrm{\etaup }}{\xi }_{\mathrm{fff}}-f^2_Rf^3_{\eta }{\xi }_{\mathrm{ffff}}-2f_{\mathrm{\etaup }\mathrm{\etaup }\mathrm{\etaup }\mathrm{R}}{\xi }_R-6f_{\mathrm{\etaup }\mathrm{R}}f_{\mathrm{\etaup }\mathrm{\etaup }}{\xi }_{\mathrm{Rf}}-6f_{\eta }f_{\mathrm{\etaup }\mathrm{\etaup }\mathrm{R}}{\xi }_{\mathrm{Rf}}\\\\
\ \ \ \ \ \ \ \ \ \ \ -2f_Rf_{\mathrm{\etaup }\mathrm{\etaup }\mathrm{\etaup }}{\xi }_{\mathrm{Rf}}-6f^2_{\eta }f_{\mathrm{\etaup }\mathrm{R}}{\xi }_{\mathrm{Rff}}-6f_Rf_{\eta }f_{\mathrm{\etaup }\mathrm{\etaup }}{\xi }_{\mathrm{Rff}}-2f_Rf^3_{\eta }{\xi }_{\mathrm{Rfff}}-f_{\mathrm{\etaup }\mathrm{\etaup }\mathrm{\etaup }}{\xi }_{\mathrm{RR}}-3f_{\eta }f_{\mathrm{\etaup }\mathrm{\etaup }}{\xi }_{\mathrm{RRf}}\\\\
\ \ \ \ \ \ \ \ \ \ \ -f^3_{\eta }{\xi }_{\mathrm{RRff}}-2f_{\mathrm{\etaup }\mathrm{\etaup }\mathrm{RR}}{\xi }_{\eta }-4f^2_{\mathrm{\etaup }\mathrm{R}}{\xi }_{\mathrm{\etaup }\mathrm{f}}-4f_{\eta }f_{\mathrm{\etaup }\mathrm{RR}}{\xi }_{\mathrm{\etaup }\mathrm{f}}-2f_{\mathrm{RR}}f_{\mathrm{\etaup }\mathrm{\etaup }}{\xi }_{\mathrm{\etaup }\mathrm{f}}-4f_Rf_{\mathrm{\etaup }\mathrm{\etaup }\mathrm{R}}{\xi }_{\mathrm{\etaup }\mathrm{f}}-2f_{\mathrm{RR}}f^2_{\eta }{\xi }_{\mathrm{\etaup }\mathrm{ff}}\\\\
\ \ \ \ \ \ \ \ \ \ \ -8f_Rf_{\eta }f_{\mathrm{\etaup }\mathrm{R}}{\xi }_{\mathrm{\etaup }\mathrm{ff}}-2f^2_Rf_{\mathrm{\etaup }\mathrm{\etaup }}{\xi }_{\mathrm{\etaup }\mathrm{ff}}-2f^2_Rf^2_{\eta }{\xi }_{\mathrm{\etaup }\mathrm{fff}}-4f_{\mathrm{\etaup }\mathrm{\etaup }\mathrm{R}}{\xi }_{\mathrm{\etaup }\mathrm{R}}-8f_{\eta }f_{\mathrm{\etaup }\mathrm{R}}{\xi }_{\mathrm{\etaup }\mathrm{Rf}}-4f_Rf_{\mathrm{\etaup }\mathrm{\etaup }}{\xi }_{\mathrm{\etaup }\mathrm{Rf}}\\\\
\ \ \ \ \ \ \ \ \ \ \ -4f_Rf^2_{\eta }{\xi }_{\mathrm{\etaup }\mathrm{Rff}}-2f_{\mathrm{\etaup }\mathrm{\etaup }}{\xi }_{\mathrm{\etaup }\mathrm{RR}}-2f^2_{\eta }{\xi }_{\mathrm{\etaup }\mathrm{RRf}}-f_{\mathrm{\etaup }\mathrm{RR}}{\xi }_{\mathrm{\etaup }\mathrm{\etaup }}-f_{\mathrm{RR}}f_{\eta }{\xi }_{\mathrm{\etaup }\mathrm{\etaup }\mathrm{f}}-2f_Rf_{\mathrm{\etaup }\mathrm{R}}{\xi }_{\mathrm{\etaup }\mathrm{\etaup }\mathrm{f}}-f^2_Rf_{\eta }{\xi }_{\mathrm{\etaup }\mathrm{\etaup }\mathrm{ff}}\\\\
\ \ \ \ \ \ \ \ \ \ \ -2f_{\mathrm{\etaup }\mathrm{R}}{\xi }_{\mathrm{\etaup }\mathrm{\etaup }\mathrm{R}}-2f_Rf_{\eta }{\xi }_{\mathrm{\etaup }\mathrm{\etaup }\mathrm{Rf}}-f_{\eta }{\xi }_{\mathrm{\etaup }\mathrm{\etaup }\mathrm{RR}},\\\\

\end{array}
\end{equation*}

\begin{equation*}
    \begin{array}{lr}
{\kappa }^{\mathrm{\etaup }\mathrm{RR}}=-3f_Rf_{\mathrm{\etaup }\mathrm{RR}}{\zeta }_f-3f^2_Rf_{\mathrm{\etaup }\mathrm{R}}{\zeta }_{\mathrm{ff}}-f^3_Rf_{\eta }{\zeta }_{\mathrm{fff}}-2f_{\mathrm{\etaup }\mathrm{RR}}{\zeta }_R-4f_Rf_{\mathrm{\etaup }\mathrm{R}}{\zeta }_{\mathrm{Rf}}-2f^2_Rf_{\eta }{\zeta }_{\mathrm{Rff}}\\\\
\ \ \ \ \ \ \ \ \ \ \ -f_{\mathrm{RRR}}(f_{\eta }{\zeta }_f+{\zeta }_{\eta })-f^3_R{\zeta }_{\mathrm{\etaup }\mathrm{ff}}-f_R{\zeta }_{\mathrm{\etaup }\mathrm{fff}}-2f^2_R{\zeta }_{\mathrm{\etaup }\mathrm{Rf}}-f_{\mathrm{\etaup }\mathrm{R}}{\zeta }_{\mathrm{\etaup }\mathrm{\etaup }\mathrm{ff}}-f_Rf_{\eta }{\zeta }_{\mathrm{\etaup }\mathrm{\etaup }\mathrm{\etaup }\mathrm{f}}+f_{\mathrm{\etaup }\mathrm{RR}}{\kappa }_f\\\\
\ \ \ \ \ \ \ \ \ \ \ + 2f_Rf_{\mathrm{\etaup }\mathrm{R}}{\kappa }_{\mathrm{ff}}+f^2_Rf_{\eta }{\kappa }_{\mathrm{fff}}+2f_{\mathrm{\etaup }\mathrm{R}}{\kappa }_{\mathrm{Rf}} + 2f_Rf_{\eta }{\kappa }_{\mathrm{Rff}} +f^2_R{\kappa }_{\mathrm{\etaup }\mathrm{ff}}+{\kappa }_{\mathrm{\etaup}\mathrm{fff}}+2f_R{\kappa }_{\mathrm{\etaup }\mathrm{Rf}}+f_{\eta }{\kappa }_{\mathrm{\etaup }\mathrm{\etaup }\mathrm{\etaup }\mathrm{f}}\\\\
\ \ \ \ \ \ \ \ \ \ \ -2f^2_{\mathrm{\etaup }\mathrm{R}}{\xi }_f-2f_{\eta }f_{\mathrm{\etaup }\mathrm{RR}}{\xi }_f-2f_Rf_{\mathrm{\etaup }\mathrm{\etaup }\mathrm{R}}{\xi }_f-4f_Rf_{\eta }f_{\mathrm{\etaup }\mathrm{R}}{\xi }_{\mathrm{ff}} -f^2_Rf_{\mathrm{\etaup }\mathrm{\etaup }}{\xi }_{\mathrm{ff}} - f^2_Rf^2_{\eta }{\xi }_{\mathrm{fff}}-2f_{\mathrm{\etaup }\mathrm{\etaup }\mathrm{R}}{\xi }_R\\\\
\ \ \ \ \ \ \ \ \ \ \ -4f_{\eta }f_{\mathrm{\etaup }\mathrm{R}}{\xi }_{\mathrm{Rf}}-2f_Rf_{\mathrm{\etaup }\mathrm{\etaup }}{\xi }_{\mathrm{Rf}} - 2f_Rf^2_{\eta }{\xi }_{\mathrm{Rff}}-f_{\mathrm{\etaup }\mathrm{RR}}{\xi }_{\eta } -2f_Rf_{\mathrm{\etaup }\mathrm{R}}{\xi }_{\mathrm{\etaup }\mathrm{f}}-f_{\mathrm{RR}}(3f_{\mathrm{\etaup }\mathrm{R}}{\zeta }_f+2f_{\eta }{\zeta }_{\mathrm{Rf}}\\\\
\ \ \ \ \ \ \ \ \ \ \ +3f_R(f_{\eta }{\zeta }_{\mathrm{ff}}+{\zeta }_{\mathrm{\etaup }\mathrm{f}})+2{\zeta }_{\mathrm{\etaup }\mathrm{R}}-f_{\eta }{\kappa }_{\mathrm{ff}}-{\kappa }_{\mathrm{\etaup }\mathrm{f}}+f_{\mathrm{\etaup }\mathrm{\etaup }}{\xi }_f +f^2_{\eta }{\xi }_{\mathrm{ff}}+f_{\eta }{\xi }_{\mathrm{\etaup }\mathrm{f}})-f^2_Rf_{\eta }{\xi }_{\mathrm{\etaup }\mathrm{ff}}-f_{\eta }{\xi }_{\mathrm{\etaup }\mathrm{fff}}\\\\
\ \ \ \ \ \ \ \ \ \ \ -2f_{\mathrm{\etaup }\mathrm{R}}{\xi }_{\mathrm{\etaup }\mathrm{R}}-2f_Rf_{\eta }{\xi }_{\mathrm{\etaup }\mathrm{Rf}}-f_{\mathrm{\etaup }\mathrm{\etaup }}{\xi }_{\mathrm{\etaup }\mathrm{\etaup }\mathrm{ff}}-f^2_{\eta }{\xi }_{\mathrm{\etaup }\mathrm{\etaup }\mathrm{\etaup }\mathrm{f}},\\\\

{\kappa }^{RR}=-f^3_R{\zeta }_{\mathrm{ff}}+{\kappa }_{\mathrm{RR}}+f_{\mathrm{RR}}\left(-2{\zeta }_R+{\kappa }_f-f_{\eta }{\xi }_f\right)+f^2_R\left(-2{\zeta }_{\mathrm{Rf}}+{\kappa }_{\mathrm{ff}}-f_{\eta }{\xi }_{\mathrm{ff}}\right)-2f_{\mathrm{\etaup }\mathrm{R}}{\xi }_R\\\\
\ \ \ \ \ \ \ \ \ \ \ -f_R\left(3f_{\mathrm{RR}}{\zeta }_f+{\zeta }_{\mathrm{RR}}-2{\kappa }_{\mathrm{Rf}}+2f_{\mathrm{\etaup }\mathrm{R}}{\xi }_f+2f_{\eta }{\xi }_{\mathrm{Rf}}\right)-f_{\eta }{\xi }_{\mathrm{RR}} \\\\

{\kappa}^{RRR}=-3f^2_{\mathrm{RR}}{\zeta }_f-f^4_R{\zeta }_{\mathrm{fff}}-3f_{\mathrm{RRR}}{\zeta }_R+f_{\mathrm{RRR}}{\kappa }_f+{\kappa }_{\mathrm{RRR}}-f_{\mathrm{RRR}}f_{\eta }{\xi }_f+f^3_R(-3{\zeta }_{\mathrm{Rff}}+{\kappa }_{\mathrm{fff}}-f_{\eta }{\xi }_{\mathrm{fff}})\\\\
\ \ \ \ \ \ \ \ \ \ \ -3f_{\mathrm{\etaup }\mathrm{RR}}{\xi }_R- 3f_{\mathrm{RR}}(2f^2_R{\zeta }_{\mathrm{ff}}+{\zeta }_{\mathrm{RR}}-{\kappa }_{\mathrm{Rf}}+f_{\mathrm{\etaup }\mathrm{R}}{\xi }_f+f_R(3{\zeta }_{\mathrm{Rf}}-{\kappa }_{\mathrm{ff}}+f_{\eta }{\xi }_{\mathrm{ff}})+f_{\eta }{\xi }_{\mathrm{Rf}})\\\\
\ \ \ \ \ \ \ \ \ \ \ -3f^2_R({\zeta }_{\mathrm{RRf}}-{\kappa }_{\mathrm{Rff}}+f_{\mathrm{\etaup }\mathrm{R}}{\xi }_{\mathrm{ff}}+f_{\eta }{\xi }_{\mathrm{Rff}})-3f_{\mathrm{\etaup }\mathrm{R}}{\xi }_{\mathrm{RR}}-f_R(4f_{\mathrm{RRR}}{\zeta }_f+{\zeta }_{\mathrm{RRR}}-3{\kappa }_{\mathrm{RRf}}\\\\
\ \ \ \ \ \ \ \ \ \ \ +3f_{\mathrm{\etaup }\mathrm{RR}}{\xi }_f+6f_{\mathrm{\etaup }\mathrm{R}}{\xi }_{\mathrm{Rf}}+3f_{\eta }{\xi }_{\mathrm{RRf}})-f_{\eta }{\xi }_{\mathrm{RRR}},\\\\

{\kappa }^{RRRR}=-f^5_R{\zeta }_{\mathrm{ffff}}-4f_{\mathrm{RRRR}}{\zeta }_R-6f_{\mathrm{RRR}}{\zeta }_{\mathrm{RR}}+f_{\mathrm{RRRR}}{\kappa }_f+4f_{\mathrm{RRR}}{\kappa }_{\mathrm{Rf}}+{\kappa }_{\mathrm{RRRR}}-f_{\mathrm{RRRR}}f_{\eta }{\xi }_f\\\\
\ \ \ \ \ \ \ \ \ \ \ -4f_{\mathrm{RRR}}f_{\mathrm{\etaup }\mathrm{R}}{\xi }_f - 3f^2_{\mathrm{RR}}(5f_R{\zeta }_{\mathrm{ff}}+4{\zeta }_{\mathrm{Rf}}-{\kappa }_{\mathrm{ff}}+f_{\eta }{\xi }_{\mathrm{ff}})+f^4_R(-4{\zeta }_{\mathrm{Rfff}}+{\kappa }_{\mathrm{ffff}}-f_{\eta }{\xi }_{\mathrm{ffff}})\\\\
\ \ \ \ \ \ \ \ \ \ \ -4f_{\mathrm{\etaup }\mathrm{RRR}}{\xi }_R-4f_{\mathrm{RRR}}f_{\eta }{\xi }_{\mathrm{Rf}}-2f^3_R(3{\zeta }_{\mathrm{RRff}}-2{\kappa }_{\mathrm{Rfff}}+2f_{\mathrm{\etaup }\mathrm{R}}{\xi }_{\mathrm{fff}}+2f_{\eta }{\xi }_{\mathrm{Rfff}})-6f_{\mathrm{\etaup }\mathrm{RR}}{\xi }_{\mathrm{RR}}\\\\
\ \ \ \ \ \ \ \ \ \ \ -2f_{\mathrm{RR}}(5f_{\mathrm{RRR}}{\zeta }_f+5f^3_R{\zeta }_{\mathrm{fff}}+2{\zeta }_{\mathrm{RRR}}-3{\kappa }_{\mathrm{RRf}}+3f_{\mathrm{\etaup }\mathrm{RR}}{\xi }_f+3f^2_R(4{\zeta }_{\mathrm{Rff}}-{\kappa }_{\mathrm{fff}}+f_{\eta }{\xi }_{\mathrm{fff}})\\\\
\ \ \ \ \ \ \ \ \ \ \ +6f_{\mathrm{\etaup }\mathrm{R}}{\xi }_{\mathrm{Rf}} + f_R(9{\zeta }_{\mathrm{RRf}}-6{\kappa }_{\mathrm{Rff}}+6f_{\mathrm{\etaup }\mathrm{R}}{\xi }_{\mathrm{ff}}+6f_{\eta }{\xi }_{\mathrm{Rff}})+3f_{\eta }{\xi }_{\mathrm{RRf}}) -2f^2_R(5f_{\mathrm{RRR}}{\zeta }_{\mathrm{ff}}+2{\zeta }_{\mathrm{RRRf}}\\\\
\ \ \ \ \ \ \ \ \ \ \ -3{\kappa }_{\mathrm{RRff}}+3f_{\mathrm{\etaup }\mathrm{RR}}{\xi }_{\mathrm{ff}}+6f_{\mathrm{\etaup }\mathrm{R}}{\xi }_{\mathrm{Rff}}+3f_{\eta }{\xi }_{\mathrm{RRff}})-4f_{\mathrm{\etaup }\mathrm{R}}{\xi }_{\mathrm{RRR}}\\\\
\ \ \ \ \ \ \ \ \ \ \ -f_R(5f_{\mathrm{RRRR}}{\zeta }_f+{\zeta }_{\mathrm{RRRR}}-4{\kappa }_{\mathrm{RRRf}}+4f_{\mathrm{\etaup }\mathrm{RRR}}{\xi }_f+4f_{\mathrm{RRR}}(4{\zeta }_{\mathrm{Rf}}-{\kappa }_{\mathrm{ff}}+f_{\eta }{\xi }_{\mathrm{ff}})+12f_{\mathrm{\etaup }\mathrm{RR}}{\xi }_{\mathrm{Rf}}\\\\
\ \ \ \ \ \ \ \ \ \ \ +12f_{\mathrm{\etaup }\mathrm{R}}{\xi }_{\mathrm{RRf}}+4f_{\eta }{\xi }_{\mathrm{RRRf}})-f_{\eta }{\xi }_{\mathrm{RRRR}},
\end{array}
\end{equation*}

\newpage

\begin{equation*}
\begin{array}{lr}

\kappa^{\eta RRR}=-f_{RRRR} f_{\eta } \zeta_{f}-4f_{RRR} f_{\eta R} \zeta_{f}-6f_{RR} f_{\eta RR} \zeta_{f}-4f_{R} f_{\eta RRR} \zeta_{f}-3f_{RR}^2 f_{\eta } \zeta_{ff}\\\\
\ \ \ \ \ \ \ \ \ \ \ -4f_{R} f_{RRR} f_{\eta } \zeta_{ff}-12f_{R} f_{RR} f_{\eta R} \zeta_{ff}-6f_{R}^2 f_{\eta RR} \zeta_{ff} -6f_{R}^2 f_{RR} f_{\eta } \zeta_{fff} - 4f_{R}^3 f_{\eta R} \zeta_{fff}\\\\
\ \ \ \ \ \ \ \ \ \ \ -f_{R}^4 f_{\eta } \zeta_{ffff}-3f_{\eta RRR} \zeta_{R}-3f_{RRR} f_{\eta } \zeta_{Rf} - 9f_{RR} f_{\eta}R\zeta_{Rf}-9f_{R} f_{\eta RR} \zeta_{Rf}-9f_{R} f_{RR} f_{\eta}\zeta_{Rff}\\\\
\ \ \ \ \ \ \ \ \ \ \ -9f_{R}^2 f_{\eta R} \zeta_{Rff} -3f_{R}^3 f_{\eta}\zeta_{Rfff}-3f_{\eta RR} \zeta_{RR}-3f_{RR} f_{\eta}\zeta_{RRf}-6f_{R} f_{\eta R} \zeta_{RRf}-3f_{R}^2 f_{\eta} \zeta_{RRff}\\\\
\ \ \ \ \ \ \ \ \ \ \ -f_{\eta} R\zeta_{RRR}-f_R f_\eta \zeta_{RRRf}-f_{RRRR} \zeta_\eta -3f_{RR}^2 \zeta_{\eta  f}-4f_R f_{RRR} \zeta_{\eta  f} -6f_{R}^2 f_{RR} \zeta_{\eta ff}-f_{R}^4 \zeta_{\eta fff}\\\\
\ \ \ \ \ \ \ \ \ \ \ -3f_{RRR} \zeta_{\eta R}-9f_R f_{RR} \zeta_{\eta Rf}-3f_{R}^3 \zeta_{\eta Rff}-3f_{RR} \zeta_{\eta RR} -3f_{R}^2 \zeta_{\eta RRf}-f_R \zeta_{\eta RRR}+ f_{\eta RRR} \kappa_f\\\\
\ \ \ \ \ \ \ \ \ \ \ +f_{RRR} f_\eta  \kappa_{ff} + 3f_{RR} f_{\eta R} \kappa_{ff} + 3f_{R} f_{\eta RR} \kappa_{ff} + 3f_{R} f_{RR} f_{\eta} \kappa_{fff} + 3f_{R}^2 f_{\eta R} \kappa_{fff} + f_{R}^3 f_\eta  \kappa_{ffff}\\\\
\ \ \ \ \ \ \ \ \ \ \ + 3f_{\eta RR} \kappa_{Rf} + 3f_{RR} f_{\eta} \kappa_{Rff} + 6f_R f_{\eta R} \kappa_{Rff} + 3f_{R}^2 f_\eta  \kappa_{Rfff} + 3f_{\eta R} \kappa_{RRf} + 3f_R f_\eta  \kappa_{RRff}\\\\
\ \ \ \ \ \ \ \ \ \ \ + f_\eta  \kappa_{RRRf} + f_{RRR} \kappa_\eta f+3f_R f_{RR} \kappa_{\eta ff}+ f_{R}^3 \kappa_{\eta fff}+ 3f_{RR} \kappa_{\eta Rf}+ 3f_{R}^2 \kappa_{\eta Rff} + 3f_{R} \kappa_{\eta RRf}\\\\
\ \ \ \ \ \ \ \ \ \ \ + \kappa_{\eta RRR}-6f_{\eta R} f_{\eta RR} \xi_f - 2f_\eta  f_{\eta RRR} \xi_f- f_{RRR} f_{\eta \eta} \xi_f - 3f_{RR} f_{\eta \eta R} \xi_f - 3f_R f_{\eta \eta RR} \xi_f\\\\
\ \ \ \ \ \ \ \ \ \ \ - f_{RRR} f_{\eta}^2 \xi_{ff} - 6f_{RR} f_\eta f_{\eta R} \xi_{ff} - 6f_R f_{\eta R}^2 \xi_{ff} - 6f_R f_\eta f_{\eta RR} \xi_{ff} - 3f_R f_{RR} f_{\eta \eta} \xi_{ff}\\\\
\ \ \ \ \ \ \ \ \ \ \ - 3f_{R}^2 f_{\eta \eta R} \xi_{ff} - 3f_R f_{RR} f_{\eta}^2 \xi_{fff} - 6f_{R}^2 f_\eta f_{\eta R} \xi_{fff}-f_{R}^3 f_{\eta \eta} \xi_{fff} - f_{R}^3 f_{\eta}^2 \xi_{ffff} - 3f_{\eta \eta RR} \xi_R\\\\
\ \ \ \ \ \ \ \ \ \ \ - 6f_{\eta R}^2 \xi_{Rf} - 6f_\eta  f_{\eta RR} \xi_{Rf} - 3f_{RR} f_{\eta \eta}\xi_{Rf} - 6f_R f_{\eta \eta R} \xi_{Rf} - 3f_{RR} f_{\eta}^2 \xi_{Rff}- 12f_R f_\eta f_{\eta R} \xi_{Rff}\\\\
\ \ \ \ \ \ \ \ \ \ \ - 3f_{R}^2 f_{\eta \eta} \xi_{Rff} - 3f_{R}^2 f_{\eta}^2 \xi_{Rfff} - 3f_{\eta \eta R} \xi_{RR} - 6f_\eta  f_{\eta R} \xi_{RRf} - 3f_R f_{\eta \eta} \xi_{RRf}- 3f_R f_{\eta}^2 \xi_{RRff}\\\\
\ \ \ \ \ \ \ \ \ \ \ - f_{\eta \eta} \xi_{RRR} - f_{\eta}^2 \xi_{RRRf} - f_{\eta RRR} \xi_{\eta} - f_{RRR} f_{\eta} \xi_{\eta f} - 3f_{RR} f_{\eta R} \xi_{\eta f} - 3f_{R} f_{\eta RR} \xi_{\eta f}\\\\
\ \ \ \ \ \ \ \ \ \ \ - 3f_R f_{RR} f_{\eta} \xi_{\eta ff} - 3f_{R}^2 f_{\eta R} \xi_{\eta ff}-f_{R}^3 f_{\eta} \xi_{\eta fff} - 3f_{\eta RR} \xi_{\eta R} - 3f_{RR} f_\eta \xi_{\eta Rf} - 6f_R f_{\eta R} \xi_{\eta Rf}\\\\
\ \ \ \ \ \ \ \ \ \ \ - 3f_{R}^2 f_\eta \xi_{\eta Rff} - 3f_{\eta R} \xi_{\eta RR} - 3f_R f_\eta \xi_{\eta RRf} - f_{\eta} \xi_{\eta RRR}
\end{array}
\end{equation*}

\bibliographystyle{jfm}
\bibliography{jfm-instructions}

\end{document}